\documentclass[final,5p,times,twocolumn]{elsarticle}
\usepackage {xcolor}

\usepackage{amssymb}

\begin{document}

\begin{frontmatter}



\title{Horizontal 1 K refrigerator with novel loading mechanism for polarized solid targets}


\author[JLab]{J. Brock}
\author[JLab]{C. Carlin}
\author[JLab]{D. Griffith}
\author[JLab]{M. Hoegerl}
\author[JLab]{P. Hood}
\author[JLab]{C. Flanagan}
\author[JLab]{T. Kageya}
\author[JLab]{C.D.~Keith}
\author[ODU,Bonn]{V.~Lagerquist}
\author[JLab]{J.D.~Maxwell}
\author[JLab]{D.M.~Meekins}
\author[ODU,MIT]{P.~Pandey}
\author[JLab]{S.~Witherspoon}

\affiliation[JLab]{organization={Thomas Jefferson National Accelerator Facility},
            city={Newport News},
            postcode={23606}, 
            state={VA},
            country={USA}}
\affiliation[ODU]{organization={Old Dominion University},
            city={Norfolk},
            postcode={23529}, 
            state={VA},
            country={USA}}
\affiliation[Bonn]{organization={Physikalsiches Institut Universit${\ddot a}$t \,Bonn},
            city={Bonn},
            country={Germany}}
\affiliation[MIT]{organization={Massachusetts Institute of Technology},
            city={Cambridge},
            postcode={02139}, 
            state={MA},
            country={USA}}
\begin{abstract}
We describe a helium evaporation refrigerator used to cool dynamically polarized proton and deuteron targets for electron-scattering experiments using the CEBAF Large Acceptance Spectrometer CLAS12 at Jefferson Lab.
The geometry of the CLAS12 detector systems places severe design and construction constraints on the refrigerator and its ancillary equipment, resulting in a horizontal cryostat with a length of 4~m.  The 16~cm\textsuperscript{3} target samples, consisting of frozen ammonia (NH\textsubscript{3} or ND\textsubscript{3}), are loaded at the upstream end of the cryostat and moved to the beam-interaction region using a novel transport mechanism.  At this location they are cooled with superfluid helium and polarized via dynamic nuclear polarization at 1~K and 5~T.
In this manner samples can be replaced and cooled to 1~K in about 30 minutes without disturbing any elements of the electron beam line or particle detection system.  We estimate that this method saved 18 days of valuable beam time over the course of a recent, 88-day long experiment.
\end{abstract}


\end{frontmatter}


\section{Introduction}
\label{sec:Introduction}
Polarized solid targets provide a compact, dense source of spin-polarized nuclear particles for scattering experiments that examine the substructure of protons and neutrons in terms of their constituent quarks and gluons. Since their inception more than five decades ago \cite{Abragam_1962,Chamberlain_1963}, polarized solid targets have been utilized at numerous laboratories throughout the world and continue to be essential tools that operate at the forefront of modern-day nuclear and particle physics.  In this article, we detail one important component---the 1~K refrigerator---of a new dynamically polarized target that is specifically designed and constructed to operate inside the detector systems of CLAS12, Jefferson Lab's 12~GeV CEBAF Large Acceptance Spectrometer~\cite{CLAS12}.  

Together, the target and detector are used in the \textit{Run Group C} suite of experiments that study the 3D distribution of quarks within protons and neutrons using deep inelastic scattering of electrons from polarized hydrogen and deuteron nuclei in samples of solid ammonia, NH$_3$ and ND$_3$.  As is often the case, the design of the target is driven by the geometry of the detector, resulting in a 4~m long horizontal cryostat that extends into the center of a 5~T superconducting magnet and narrows to a diameter of only 10~cm at the beam-target interaction region.  The ammonia target samples are cooled to 1~K by immersion in a shallow bath of superfluid helium and exposed to approximately 1~W of 140~GHz microwaves to drive the dynamic nuclear polarization process.  Proton (H) and deuteron (D) polarizations exceeding 90\% and 50\%, respectively, have been achieved under these conditions.  

The samples must be removed from the bath every two or three days and warmed to approximately 100~K to repair radiation damage inflicted by the incident electron beam.  Additionally, unpolarized samples of carbon and polyethylene are periodically used to mimic and quantify the scattering of electrons from nuclei other than hydrogen or deuterium.  To reduce the experimental time lost due to these configuration changes, a system was designed and implemented to replace the samples as rapidly as possible and without disassembly of any beam line elements, such as the focusing magnets and beam diagnostic monitors that are positioned directly upstream\footnote{In this article, ``downstream'' and ``upstream'' refer to the directions along and opposite the path of the electron beam, respectively.} from the target cryostat.  

The remainder of this article is organized as follows.
A brief description of the CLAS12 detector and the target cart is given in Sec.~\ref{sec:CLAS12 and the Target Cart}.  An overview of dynamic nuclear polarization is given in  Sec.~\ref{sec:DNP} to explicate the design motivations for the refrigerator and sample loading mechanism, which are detailed in Sec.~\ref{sec:Fridge} and ~\ref{sec:TargetBath}, respectively.  Their performances are presented in Sec.~\ref{sec:Performance}, and a summary is given in Sec.~\ref{sec:Summary}.

\section{CLAS12 and the Target Cart}
\label{sec:CLAS12 and the Target Cart}
\begin{figure}[t]
    \centering
    \includegraphics[width=\columnwidth]{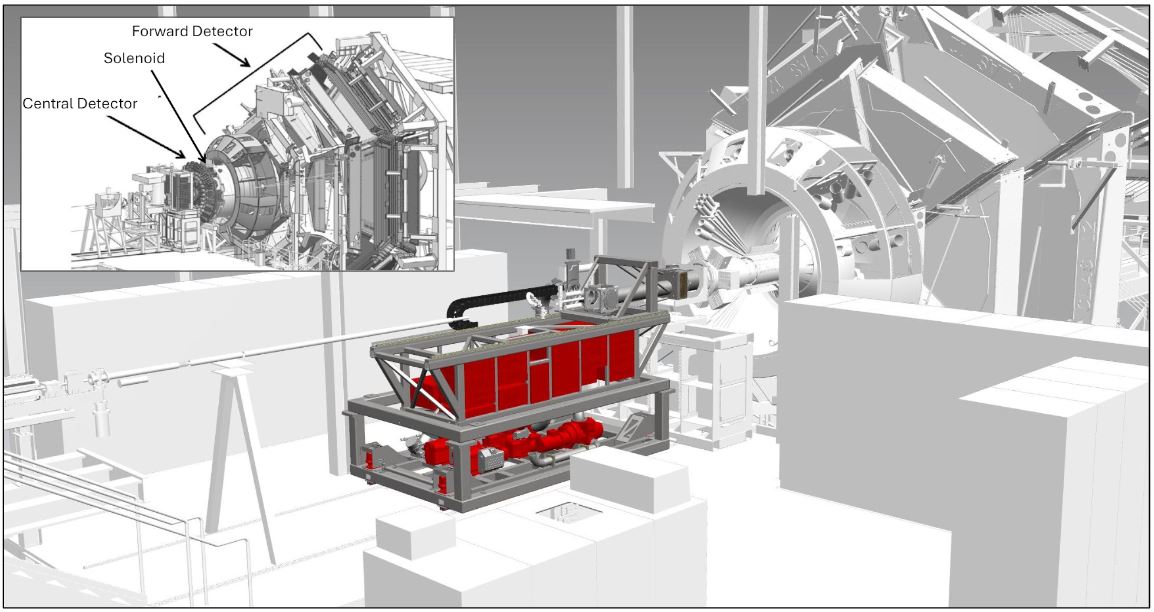}\\
    \caption{Digital rendering, with inset, of the Polarized Target installed in the Hall B CLAS12 detector at Jefferson Lab.  The electron beam traverses left to right in this figure.}
    \label{fig:CLAS12}
\end{figure}

CLAS12 is the centerpiece of the physics program in Hall B, one of four experimental end stations at Jefferson Lab.  It comprises two major detector packages, Central and Forward (inset, Fig.~\ref{fig:CLAS12}). Each features a large, superconducting magnet~\cite{Fair_2020} to aid in tracking and identification of scattered particles.  A six-coil torus magnet is used for the Forward detector, while a 5~T solenoid is part of the Central detector. The solenoid serves multiple functions for CLAS12 and is particularly important for the polarized target.  First, it bends low-energy (0.3--1.5~GeV) charged particles for spectroscopic analysis by the various elements of the Central Detector. Second, it focuses the copious number of electrons scattered by the M{\o}ller process away from the Forward Detector, thus allowing it to operate at higher data rates.  Third, the solenoid provides the magnetic field required to dynamically polarize the target in the manner described in the following section.  

\begin{figure}[t]
    \centering
    \includegraphics[width=0.8\columnwidth]{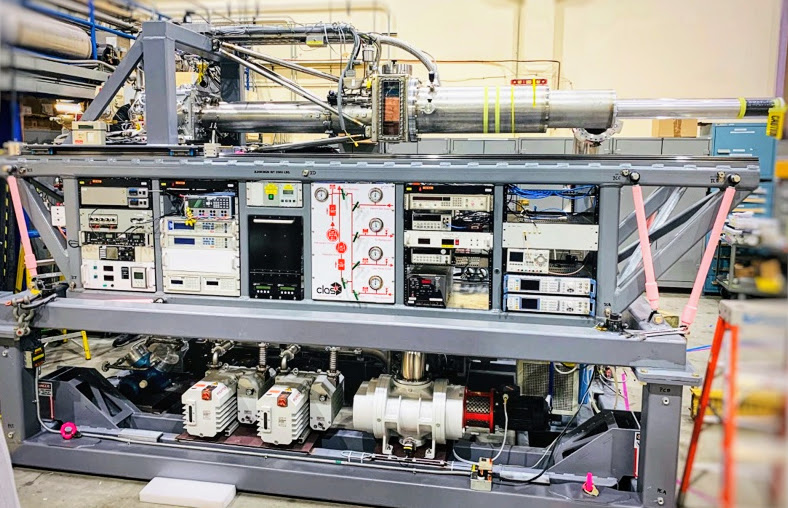}\\
    \includegraphics[width=0.8\columnwidth]{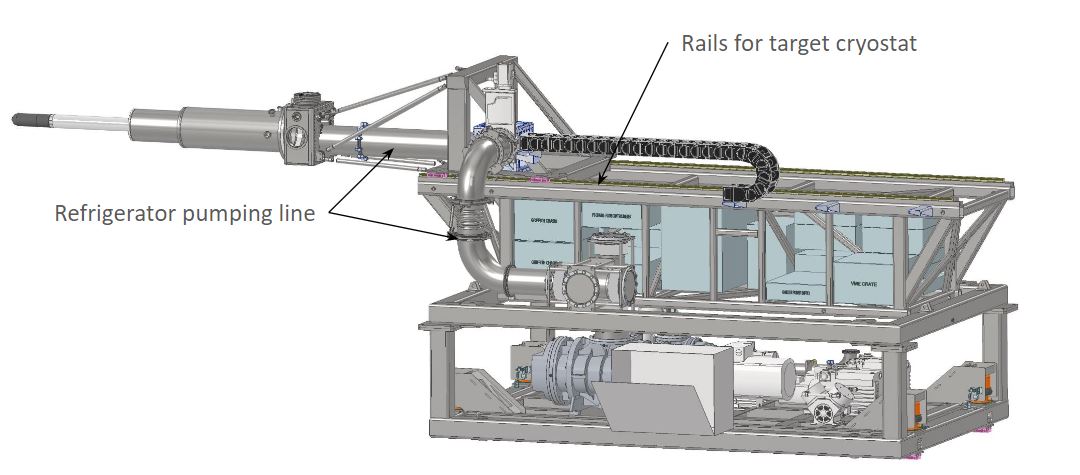}\\
    \includegraphics[width=\columnwidth]{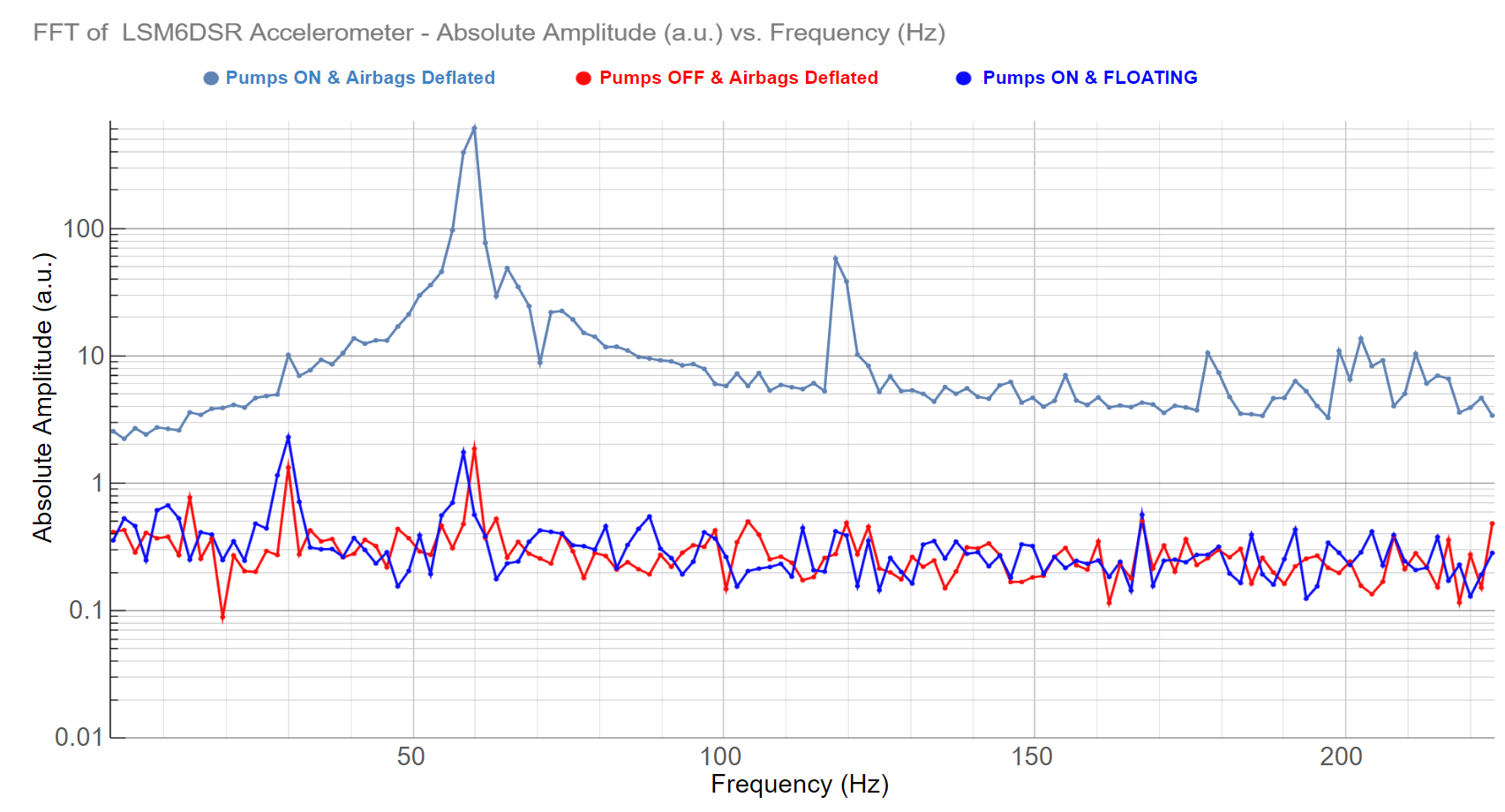}
    \caption{Top: Photograph of the target assembly, retracted, from the beam-right side, prior to transport to Hall B.
     Middle: The upper tier has a separate set of rails to extend the cryostat into the center of the CLAS12 detector.  Bottom: Fast Fourier Transform (FFT) of the measured vibration at the target center position under various conditions with the refrigerator fully extended. The resonance from the 60~Hz pump motors is clearly identifiable.}
    \label{fig:PolTarCart}
\end{figure}

The nominal location of targets within CLAS12 is at the center of the solenoid, which has a warm bore 1.8~m long and 0.78~m in diameter. Much of this volume is occupied by three concentric detector packages: the Central Neutron Detector, Central Time-of-Flight Detector, and Silicon Vertex Tracker.  The remaining space for the target is confined to a 10~cm diameter cylindrical volume concentric with the electron beam. Moreover, control electronics for detector elements are placed directly upstream of the solenoid, further complicating the design and placement of the polarized target's cryostat and ancillary equipment.

To satisfy these constraints and to facilitate easy installation and alignment in the experimental hall,
all components of the polarized target are mounted onto a three-tiered cart that can be inserted into and retracted from the Central Detector via a set of rails parallel to the electron beam line (Fig.~\ref{fig:PolTarCart}).
The nine vacuum pumps necessary to operate the target are mounted to a subframe of the bottommost tier that is vibration-isolated from the system by self-leveling airbags.  Electronics, along with a small gas handling panel, are mounted to the middle tier of the cart, which includes a second set of linear rails at its top.  A support frame for the target cryostat rolls along these rails, as the cryostat must be cantilevered 2~m beyond the cart in order to reach the center of the solenoid.  The cryostat is supported from this frame by six independently adjustable tie rods, giving six degrees of freedom and providing sub-millimeter alignment inside the Central Detector.  Three orthogonal pairs of edge-welded bellows in the pumping line between the cryostat and vacuum pumps provide further vibration isolation.  Together with the airbags, vibration from the pumps is reduced more than two orders of magnitude (Fig.~\ref{fig:PolTarCart}).

\section{Dynamic Nuclear Polarization}
\label{sec:DNP}
The most basic method by which nuclear spins can be polarized is the so-called ``static'' or ``brute-force'' method, whereby the sample is cooled to a low temperature and exposed to a strong magnetic field.  The resulting populations of the nuclear Zeeman states correspond to a spin polarization described by the Brillouin function,
\begin{equation}
	P_{TE} = \frac{2J+1}{2J} \coth \left( \frac{2J+1}{2J}\, x\right) 
	-\frac{1}{2J} \coth \left(\frac{1}{2J}\, x \right). \label{eqn:Brillioun}
\end{equation}
Here $x \equiv \mu B/kT$, with $\mu$ being the magnetic dipole moment of the spin-$J$ particle, 
$B$ the magnetic field strength, and $T$ the lattice temperature.
For spin-1/2 particles like electrons and protons, the above equation simplifies to
\begin{equation}
         P_{TE} = \tanh \left(x \right). \label{eqn:spin1/2}
\end{equation}
The subscript {\em TE\,} specifies this as the polarization of the spins when they are in thermal equilibrium with the
solid lattice.  Because nuclear moments are small, appreciable polarizations can only
be obtained at extremely low temperatures and high magnetic fields.  In the 5~T field of the CLAS12 solenoid, the sample must be cooled below 0.01~K to achieve a proton polarization of 50\%, while the polarization of deuterons is only 10\% under the same conditions.  The ultra-low temperature requirement for brute-force polarized targets limits their use to very low intensity beams and has greatly restricted their implementation.

A more common and powerful method is dynamic nuclear polarization, or DNP~\cite{Maly_2008}.
Here, a low density ($10^{19}$--$10^{20}$~cm$^{-3}$) of paramagnetic radicals is added to the sample, either by chemical doping or by exposure to ionizing radiation.  The unpaired electrons associated with these radicals can be polarized at significantly warmer temperatures due to their larger magnetic moment: in the same 5~T field, the electron polarization is greater than 99\% at 1~K.  Microwaves near the electron spin resonance (ESR) frequency (140~GHz at 5~T) stimulate electron spin flip transitions and partially saturate the ESR line.  Cross relaxation between the electrons and nearby nuclear spins, enabled by the dipolar interaction, partially transfer the electron's initial polarization to the nuclei.  This polarization can be made parallel or anti-parallel to the magnetic field 
by choosing a microwave frequency that is slightly below or above the ESR frequency.
Because they operate at higher temperatures, dynamically polarized targets can be used with far more intense particle beams than their brute-force counterparts.  

DNP is only effective in a limited selection of materials suitable for particle scattering experiments, and those that can be polarized are evaluated according to three criteria: the fraction of nucleons in the material that are polarized, their maximum achievable polarization, and the material's ability to retain this polarization when exposed to ionizing radiation \cite{Goertz_2002}.  Frozen ammonia (NH$_3$ and ND$_3$) polarized at 1~K and 5~T has become the {\em de facto\/} standard for polarized targets in highly intense beams (up to $10^{11}$~s$^{-1}$~cm$^{-2}$) because it satisfies all three criteria reasonably well~\cite{Averett_1999, Keith_2003, Pierce_2014}.  The preparation and use of the ammonia samples is described in detail by Meyer~\cite{Meyer_2004}.  Briefly, the gas is frozen at temperatures below 195~K and crushed to small millimeter-sized granules for more efficient heat transfer at 1~K.  The granules are then irradiated under liquid argon with an electron beam to a fluence of about $10^{17}$~e$^-$~cm$^{-2}$ to produce the desired concentration of paramagnetic radicals.  Thus prepared, the samples can be stored indefinitely in liquid nitrogen until needed in the scattering experiment.  

More than one species of paramagnetic radicals are produced by the electron beam, but the amino radicals \.NH$_2$ and \.ND$_2$ have been identified as those responsible for the DNP process.  These are not stable at temperatures much above 100~K, so it is critical that samples do not warm above this temperature as they are loaded into the target cryostat.  Additional radicals such as atomic hydrogen are produced when the sample is exposed to the electron beam at 1 K, and these prove deleterious to the polarization.  This radiation damage becomes problematic after an exposure of approximately $5\cdot10^{15}$~e$^-$~cm$^{-2}$ but can be partially repaired by annealing the sample to about 90~K for several minutes.  In the target described here, the samples are annealed in a separate bath of 87~K liquid argon after approximately 48~h of beam time.  This process becomes progressively less effective, and the samples are replaced after five or six anneals.

\section{1~K Refrigerator}
\label{sec:Fridge}
\subsection{Overview}
\begin{figure*}[t]
    \centering
    \includegraphics[width=0.9\textwidth]{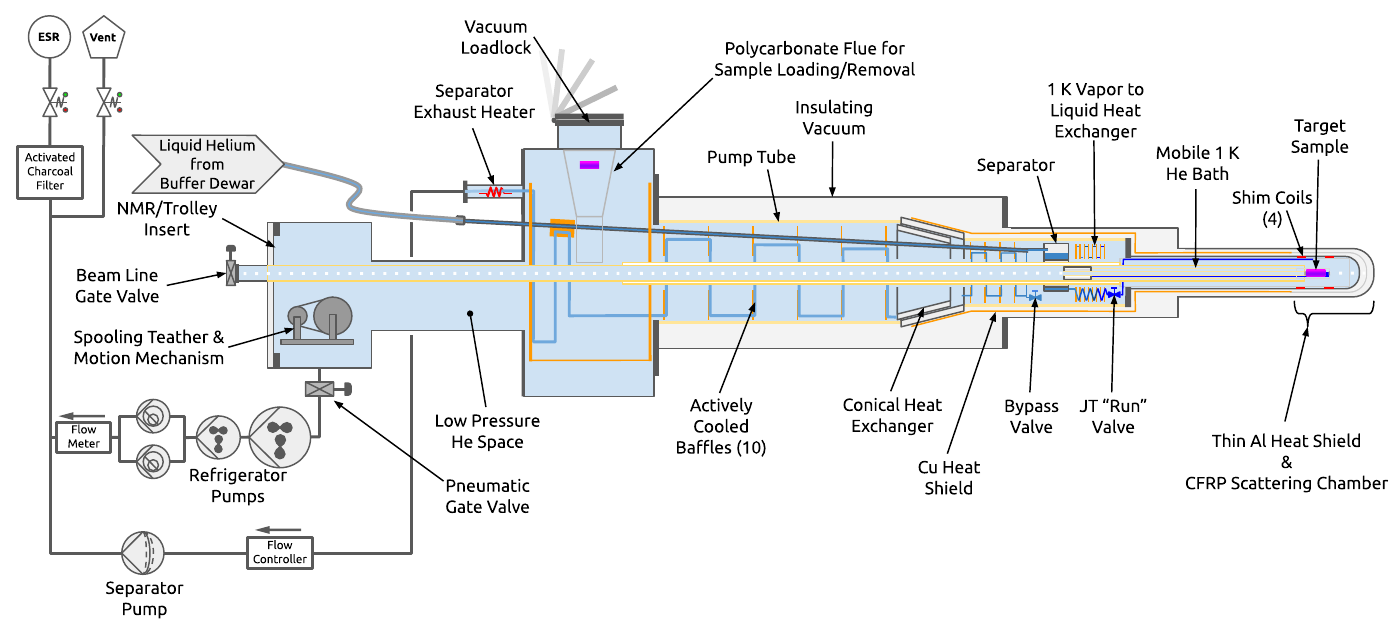}
    \caption{Simplified helium-flow diagram for the refrigerator.}
    \label{fig:FlowDiagram}
\end{figure*}

\begin{figure*}
    \centering
    \includegraphics[height=0.9\textheight]{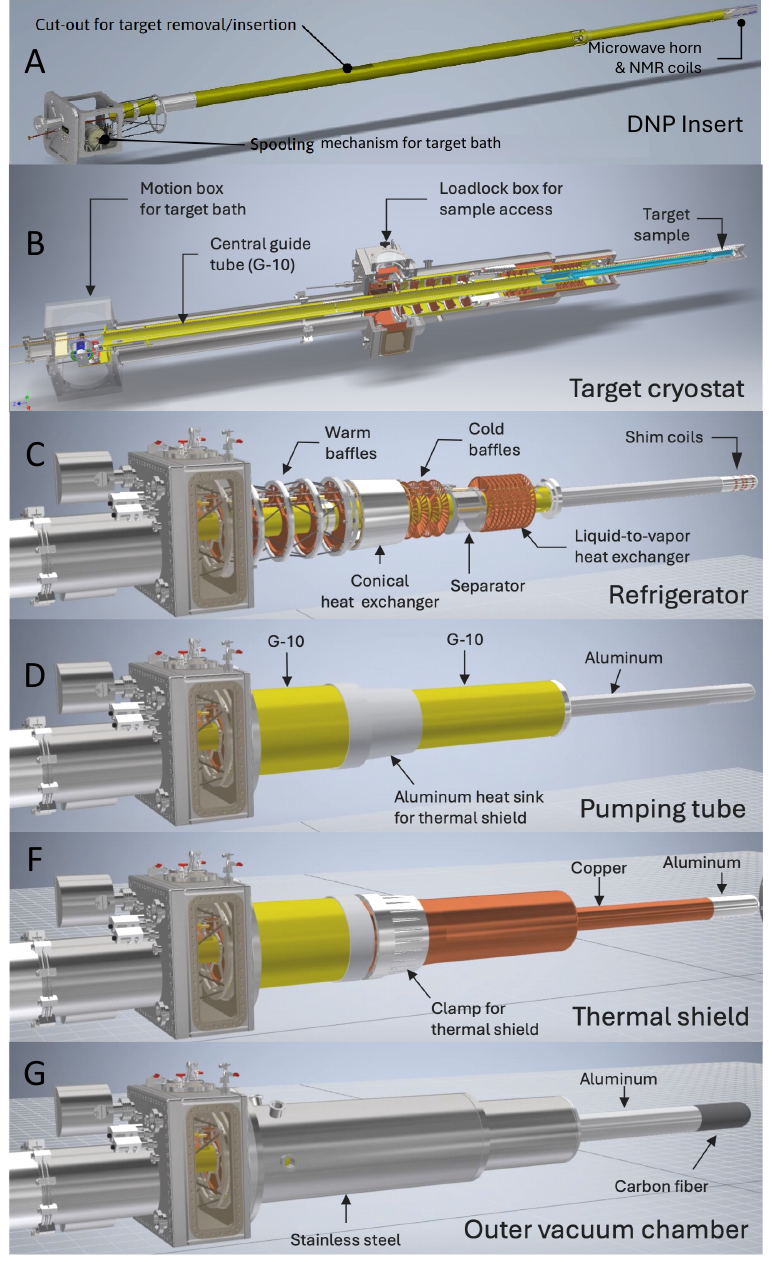}
    \caption{Digital rendering of the target cryostat showing successive layers of the system as the images progress downwards.  The overall length of the system is approximately 4.2 m. The electron beam traverses from left to right in these figures.}
    \label{fig:Layers}
\end{figure*}

The target consists of three nested elements: a 1~K refrigerator, a DNP insert housing the NMR and microwave components, and a mobile 1~K bath containing the target material. The latter two are described later in this text.   The horizontal 1~K refrigerator borrows elements from a design described by Roubeau~\cite{Roubeau_1966} and used elsewhere in vertical or near-vertical configurations~\cite{Averett_1999, Keith_2003, Pierce_2014}.  A simplified flow diagram is shown in Fig.~\ref{fig:FlowDiagram} and an annotated center-out layered rendering in Fig.~\ref{fig:Layers}.  Briefly, liquid helium is continuously siphoned from an external Dewar through a vacuum-insulated transfer line into a phase separator.  The vapor fraction of the transferred helium is pumped from the top of the separator and flows through a heat exchanger used to cool a thermal shield surrounding the refrigerator as well as a series of radiation baffles upstream of the separator.

Liquid helium from the base of the separator is used to fill an open-top bath that houses the polarized target sample. The position of this bath is not fixed inside the cryostat.  It can be moved from the in-beam location at the center of the CLAS12 solenoid to a second location about 2.5~m upstream, where the sample can be accessed and easily removed from the cryostat via a vacuum load-lock and extraction device.
Liquid delivered to the sample bath is metered by a miniature Joule-Thomson (JT) valve and pumped to a vapor pressure of about 16~Pa (corresponding to a temperature of 1~K) by a set of high-capacity Roots pumps. These pumps have a rated pumping speed of 6000~m$^3$/h for nitrogen and a measured speed of about 4500~m$^3$/h for helium. A liquid-to-vapor heat exchanger cools the liquid to approximately 2~K before it expands across the JT valve.  For faster cooling, e.g.\ during the initial cool down from room temperature and recovery of the refrigerator after changing targets, liquid from the transfer line can flow directly to the bath using a second needle valve which bypasses both the separator and the liquid-vapor heat exchanger.

The construction of the refrigerator is centered around a thin-walled G-10 tube
that serves as a guide for the removable DNP insert and mobile 1~K bath (Sec.~\ref{sec:TargetBath}).
A series of aluminum and high modulus carbon fiber truss rings make up the structure that supports the refrigerator from a stainless steel plate at the upstream end of the cryostat. Shifting the load-bearing components radially outward increases the structure's area moment of inertia and reduces the mass of support material compared to a central supporting tube of equivalent rigidity. This reduces the overall thermal conductance through the structure.

The refrigerator slides inside a larger G-10 and aluminum tube through which the bath vapor is pumped.  This pumping tube is surrounded by thermal shields made of copper and aluminum all of which are enclosed by a vacuum chamber of stainless steel, aluminum, and carbon-fiber-reinforced-polymer (CFRP) composite.  To reduce the energy loss of particles scattered from the target, all components around the beam-target interaction region are intentionally constructed from low-density materials such as aluminum and CFRP.  A series of small, superconducting
shim coils are included inside the refrigerator and
used to improve the homogeneity of the CLAS12 solenoidal
field for optimal dynamic polarization.

\begin{figure}[t]
    \centering
    \includegraphics[width=\columnwidth]{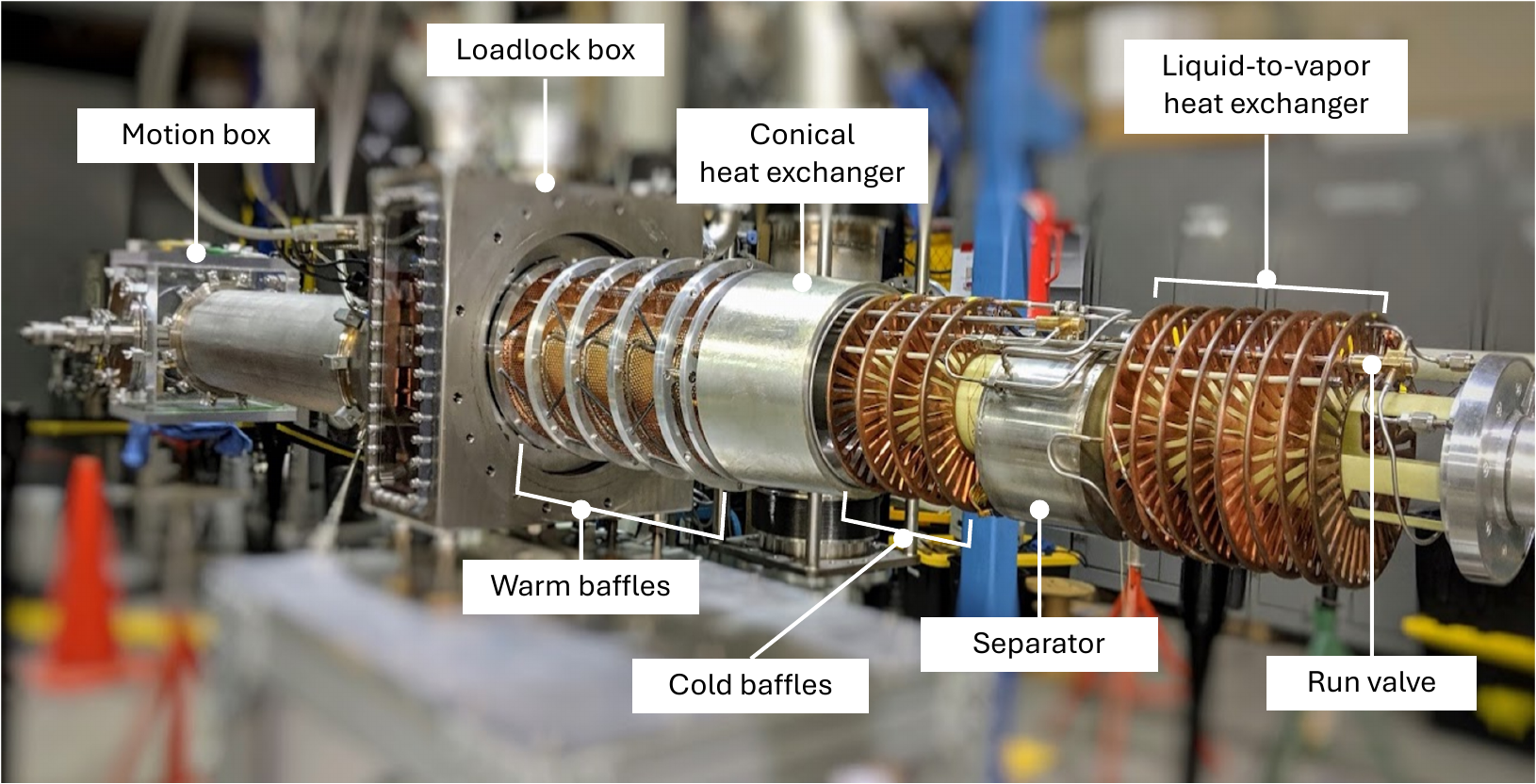}
    \caption{Annotated photograph of some of the major components of the 1~K refrigerator.  See text for details.}
    \label{fig:FridgePhoto}
\end{figure}

Individual refrigerator components are described in more detail below.  An annotated photograph of some components are 
shown in Fig.~\ref{fig:FridgePhoto}.

\subsection{Helium Dewar and Transfer Line}
\label{sec:TransferLine}
During tests of the polarized target in our fabrication facility, liquid helium was siphoned from commercial 250~L Dewars.  In experimental Hall B, liquid is drawn from a permanent 500~L Dewar that is continuously replenished by JLab's End Station Refrigerator (ESR).  The Dewar is approximately 4 meters from the refrigerator, which permits the use of a short, flexible transfer line to the phase separator. The siphon is created by pumping on the separator using a 100~L/min diaphragm pump, while a mass flow controller at the pump inlet determines the siphon rate.  The quality of the transfer line greatly influences the refrigerator performance.  Initial tests were made using a commercially-fabricated line with a 1/4-inch corrugated inner hose, but this could not provide helium to the refrigerator at liquid flow rates less than about 5 L/h without significant vaporization, resulting in unacceptable flow oscillations.  It was replaced by a higher-efficiency version constructed in-house with a non-corrugated, 1/8-inch inner tube wrapped with twenty layers of aluminized-mylar/polyester insulation with PTFE standoffs. No oscillations have been observed with this line using liquid flows as low as 1~L/h.  The transfer line feeds into the separator via a bayonet with a slight, downward angle of about 3$^{\circ}$.  

All helium pumped from the phase separator and superfluid bath are filtered through activated charcoal
and returned to the ESR as room temperature gas.  Alternatively, the gas may be vented to atmosphere outside Hall B.  
The latter path is only used during the initial cleaning of the system prior to cooling or for the first minutes 
after a target sample is loaded into the refrigerator.

\subsection{Phase Separator}
\label{sec:Separator}
The separator is an eccentric, annular stainless-steel vessel 6 inches in diameter and 4 inches long, with an offset bore for the central G-10 tube.  The upper and lower portions are separated by dense, stainless-steel wire cloth.  Liquid helium delivered to the upper part of the vessel drains through the cloth and is collected in the lower half.  The cloth also filters particulates from the helium that might otherwise plug the small diameter tubes and valves downstream from the separator.  The temperatures of the upper and lower portions of the separator's interior, as well as other refrigerator components, are measured with Cernox\textsuperscript{\small{\textregistered}}  thermometers. 

Helium gas is pumped from the upper portion of the separator through stainless steel and copper tubing and cools a series of copper baffles inside the helium pump tube, as well as the thermal shield surrounding it.  Before exiting the cryostat, the gas is warmed to room temperature by a 100~W cartridge heater inside the separator pumping line.  This prevents water from condensing on the outside of the cryostat which could then drip onto sensitive detector elements or electronics.  

\subsection{Conical Heat Exchanger and Baffles}
\label{sec:Baffles}
Copper baffles are distributed inside the pumping tube to reduce heat transport from warmer to colder components of the refrigerator due to conduction and convection of the low pressure helium vapor.  Two types of baffles are used, referred to as cold and warm. These are located on either side of a heat sink used to cool the refrigerator's thermal
shield (Sec.~\ref{sec:ThermalShield}).  
The cold baffles are fabricated with louvered openings from stamped, annealed C101 copper sheet (7.46 inches in diameter, 0.025 inches thick). The warm baffles, measuring 9.1 inches in diameter, are made from perforated C110 copper sheet (0.025 inches thick) with 1/16-inch diameter perforations spaced 3/32-inches center-to-center. The baffles are cooled by helium gas  pumped from the separator through copper tubing soldered into stamped channels on the surface of each baffle. The channels increase the contact area and minimize the solder thickness between the copper tubing and copper sheets. Brass couplers act as thermal breaks between successive baffles.  

\begin{figure}[t]
    \centering
    \includegraphics[width=\columnwidth]{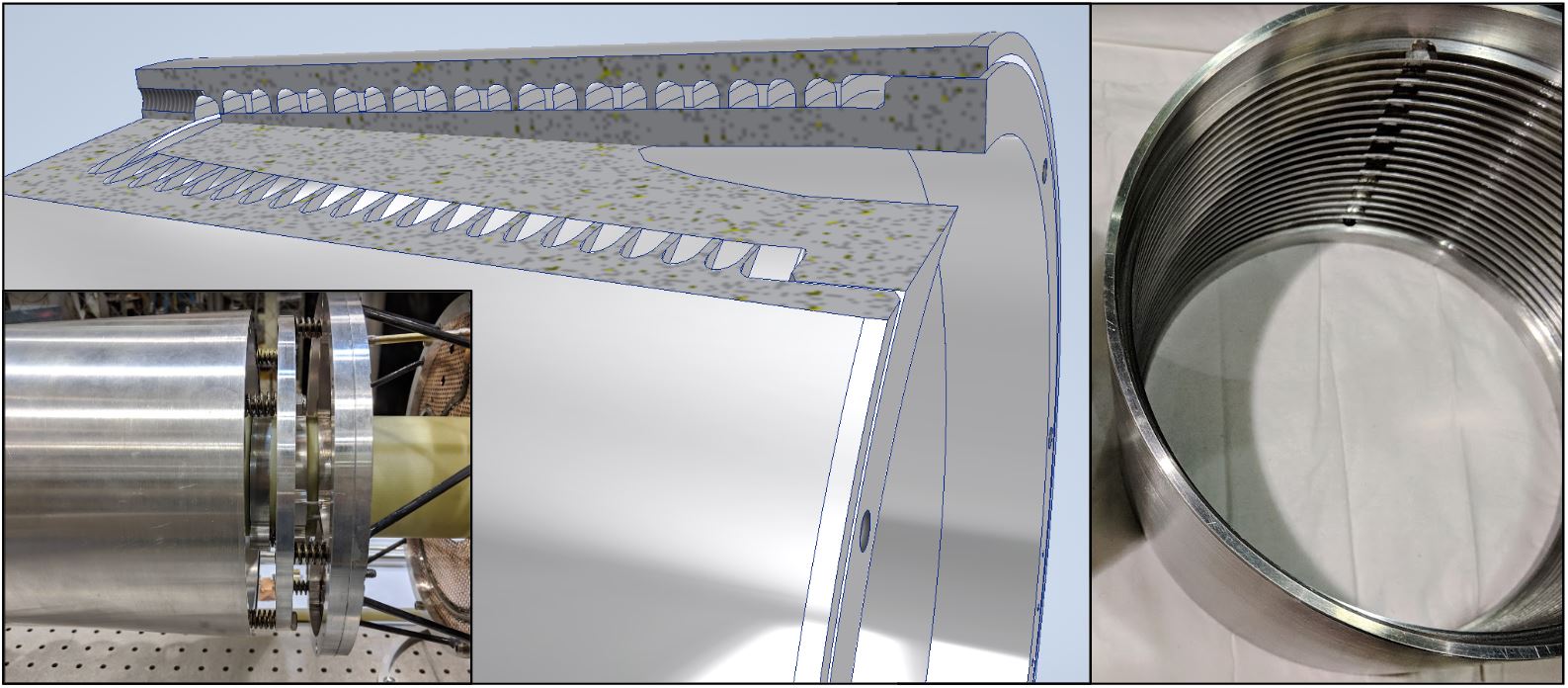}
    \caption{Sectioned illustration of the conical heat exchanger used to cool the refrigerator thermal shield. Left Inset:Spring mount of truss frame. Right Inset: Grooved flow path.}
    \label{fig:ConicalHX}
\end{figure}

A conically-shaped heat exchanger is located between the cold and warm baffles and is used to cool the outer thermal shield surrounding the low-pressure pumping tube (Sec.~\ref{sec:PumpingTube}). Silicone bronze springs mount the heat exchanger to the truss frame. Upon assembly this heat exchanger contacts a matching aluminum heat sink incorporated in the pump tube (Fig.~\ref{fig:Layers}, panel D). The conical geometry and spring compression ensures self-alignment and maintains a constant 100~N force across all temperatures, ensuring good thermal contact.  The conical heat exchanger is fabricated from two, tightly fitting shells of Al6063 as shown in Fig.~\ref{fig:ConicalHX}. A smooth inner and ribbed outer aluminum shell creates a flow space that has an equivalent diameter of 5.1~mm. 
Helium flows through the rib structure via alternating
grooves that create a series of parallel switchbacks with a total path length of 6.1~m.  This configuration provides a turbulent flow of helium coolant (Reynolds number $Re>4000$) for good heat transfer while still maintaining an acceptably low pressure drop. The shells are epoxied together using Scotch-Weld DP190 gray\footnote{3M Company.}. We have found this inexpensive, two-part epoxy to reliably bond numerous materials, including aluminum, polyimide, polyamide-imide, fiberglass G-10, and PEEK, from cryogenic to room temperatures.

\subsection{Low Temperature Heat Exchanger}
\label{HeatExchanger}
Liquid helium from the lower portion of the separator flows through a liquid-to-vapor heat exchanger where heat is transferred to the gas pumped from the 1~K superfluid bath. This heat exchanger consists of a 1/8-inch C101 copper tube soldered to a series of eight stamped and louvered C101 sheets.  A solder alloy with a superconducting transition temperature less than 1.3~K is used~\cite{Warren_1968}.\footnote{78.4 Cd/5.0 Ag/ 16.6 Zn}
     
\begin{figure}
    \centering
    \includegraphics[width=0.8\columnwidth]{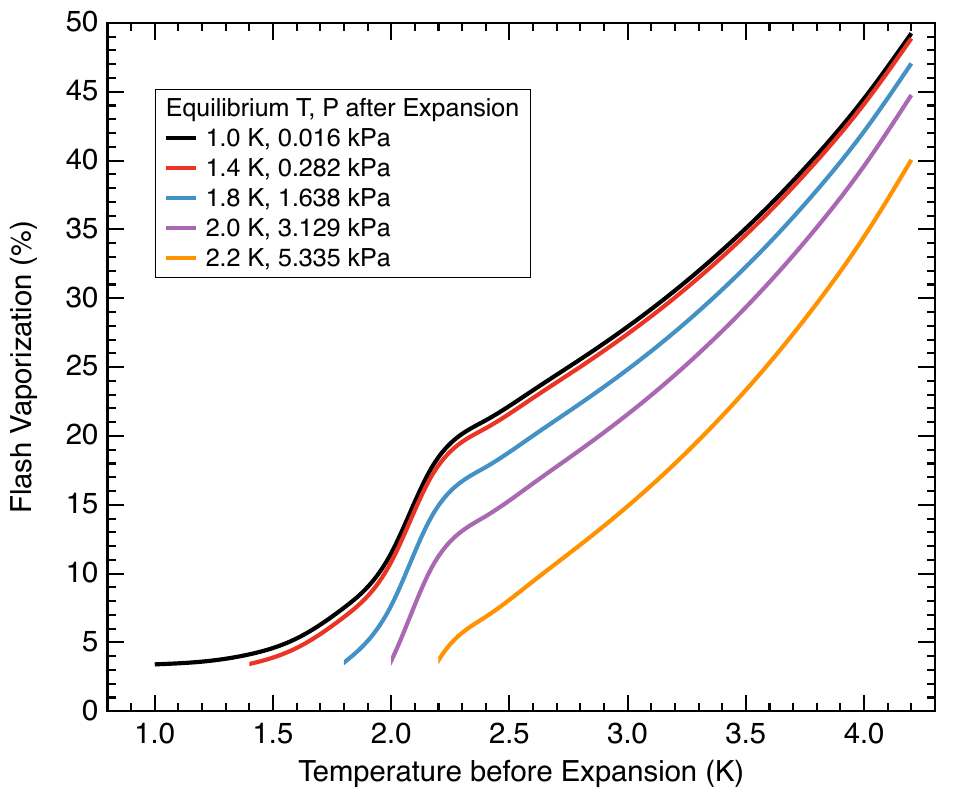}
    \caption{Percentage of $^4$He in the vapor phase following isoenthalpic expansion of 0.1~MPa liquid
    as a function of temperature prior to expansion.  The equilibrium temperature and pressure of the fluid after expansion are indicated.}
    \label{fig:FlashVapor}
\end{figure}

 The purpose of this heat exchanger is to lower the liquid's temperature before it isoenthalpically expands across a miniature Joule-Thomson valve, thus reducing flash vaporization and increasing the fraction of cold liquid delivered to the bath. Assuming the helium enters the valve entirely in the liquid phase, the fraction that vaporizes after expansion is
\begin{equation} \label{eqn:FlashVapor}
    f_V = \frac{H^L_U - H^L_D}{H^V_D - H^L_D}
\end{equation}
where $H^{L,V}_{U,D}$ is the enthalpy of the liquid (L) or vapor (V) phase evaluated at the temperature and pressure values upstream (U) and downstream (D) of the valve.  This is plotted as a function of liquid temperature upstream of the JT valve in Fig.~\ref{fig:FlashVapor}, assuming an upstream pressure of 0.1~MPa.  As indicated from the figure, subcooling the liquid in our 1~K refrigerator from 4.2~K to 1.5~K reduces the vapor content of the expanded fluid from about 50\% to less than 5\%, but there is little benefit to cooling below 1.5~K.  However, it is difficult to achieve a very high efficiency in this heat exchanger without great effect to the temperature of the target bath. Only a very modest pressure drop across the vapor side of the heat exchanger can be tolerated, and it must clearly be less than 16~Pa to achieve a bath temperature of 1.0~K.

\subsection{Needle Valves}
\label{sec:NeedleValves}
\begin{figure}[t]
    \centering
    \includegraphics[width=0.8\columnwidth]{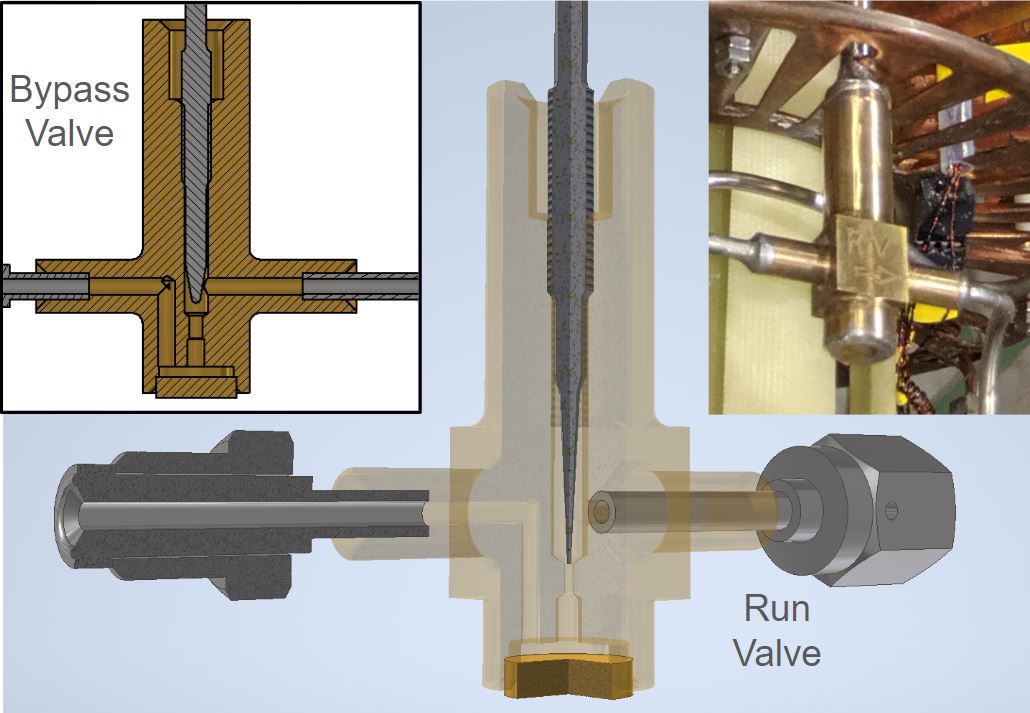}
    \includegraphics[width=0.8\columnwidth]{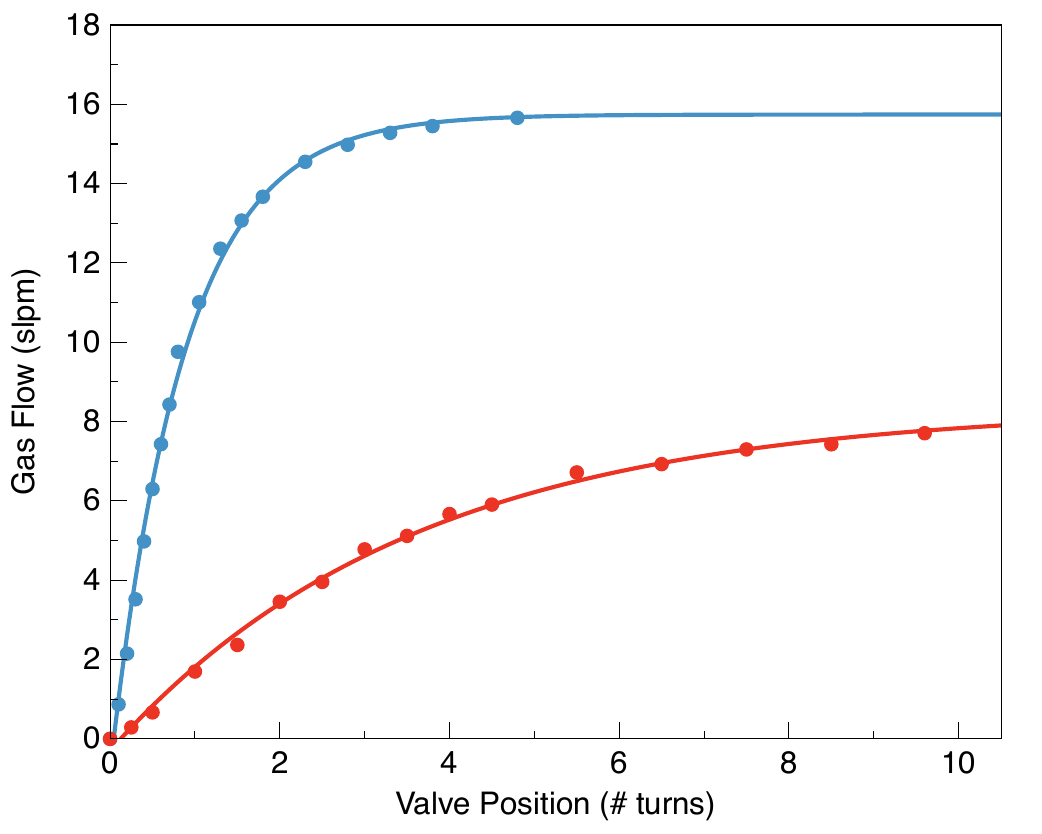}
    \caption{Top: Sectional view of the Run valve, 3$^{\circ}$ taper.
    Inset Left:Design of the Bypass valve is similar, with a coarser,
    20$^{\circ}$ taper.  Inset Right: Photograph of the Run valve.  Bottom:
    Flow of room temperature helium gas at 1.2 bar pressure differential through the Run (red) and Bypass (blue) valves.  The red and blue curves are guides to the eye.}
    \label{fig:NeedleValves}
\end{figure}

Two miniature Joule-Thomson valves control the flow of liquid helium to the sample bath.  Both feature stainless steel needles and brass bodies (Fig.~\ref{fig:NeedleValves}).  The run valve (so called because it is open when the refrigerator is running under normal conditions) meters the flow of liquid from the bottom of the separator and cooled below 2~K by the liquid-to-vapor heat exchanger (see Fig.~\ref{fig:FlowDiagram}).  The bypass valve takes liquid directly from the transfer line, thus bypassing both the separator and heat exchanger.  This valve is utilized for rapidly cooling the refrigerator from room temperature or following insertion of a target sample.

The valves are remotely controlled by stepper motors located outside the cryostat, with rotary encoders used to measure the valve position with a resolution of about  0.3$^\circ$. 
Both valves can be controlled manually (the user specifies the valve position), or by a PID feedback loop with one of the refrigerator's many process variables.  For example, a liquid level meter in the target bath is normally used to control the run valve and maintain a constant bath level, while the flow meter behind the refrigerator Roots pumps controls the bypass valve during rapid cooling.

\subsection{Shim Coils}
\label{sec:Shims}
\begin{figure}[t]
    \centering
    \includegraphics[width=0.8\columnwidth]{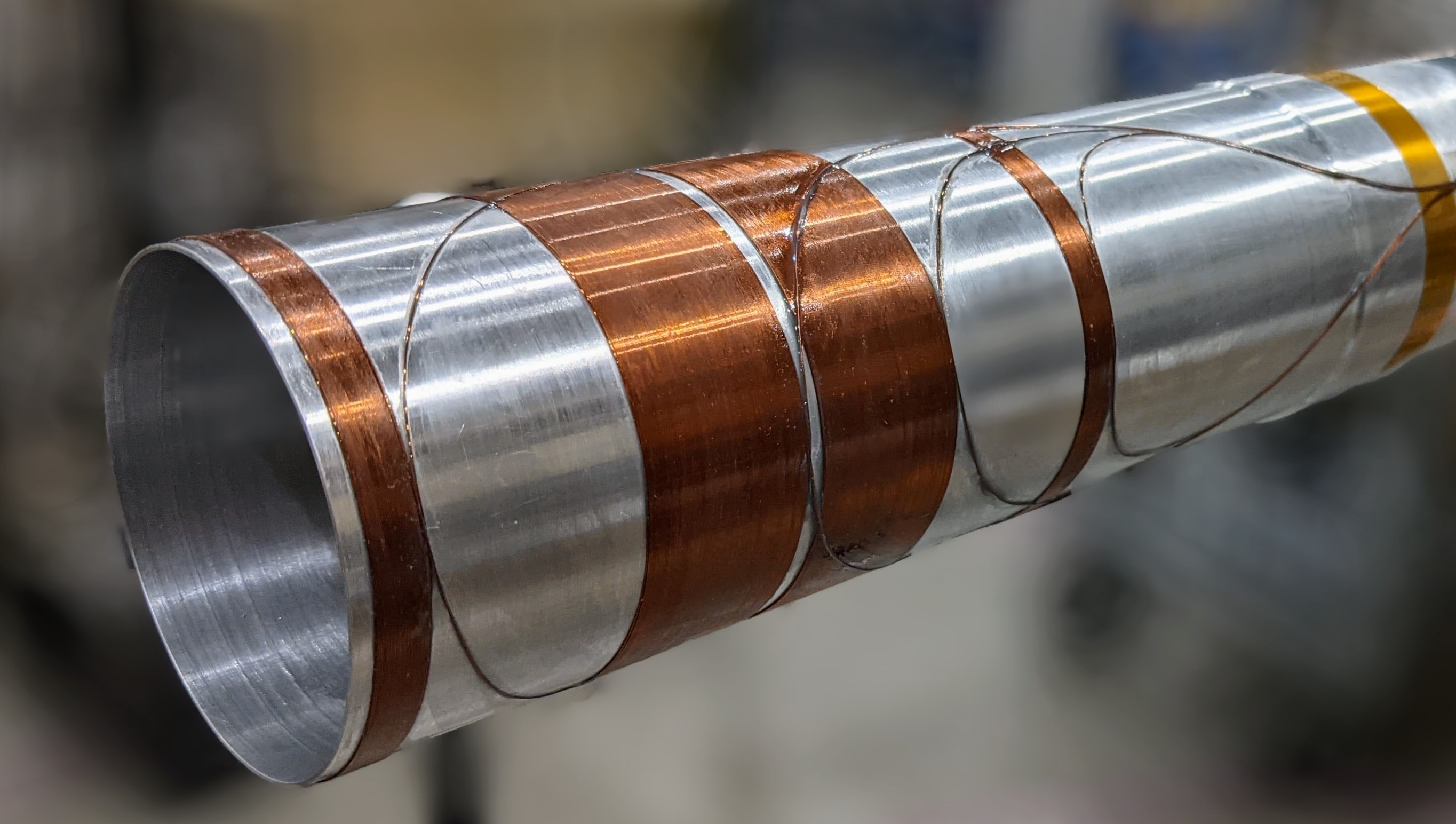}
    \caption{Superconducting shims, comprising four coils, each with two layers of windings. The center two coils each have a width of 24.5 mm (268 total windings) and are separated from each other by a 2.5 mm gap. The outer two each have a width of 6.5 mm (69 total winding) and are separated from the inner two by a gap of 28.5 mm.  The electron beam transverses right to left in this photograph.}
    \label{fig:Shims}
\end{figure}
Although the CLAS12 solenoid was designed to provide a field uniformity of $\Delta B/B \le 10^{-4}$ over the volume the target samples (2~cm diameter x 5~cm long), initial field maps indicated the actual uniformity could be up to three times worse.  Four small superconducting coils (Fig.~\ref{fig:Shims}) are therefore included inside the 1~K fridge to shim the field at the target bath~\cite{LagerquistPhD}.  The coils are wound from 0.14~mm copper-clad, multifilament NbTi wire\footnote{Supercon, Inc. type 54S43} around an aluminum mandrel, 67~mm diameter and 0.6~mm thick.  The mandrel, which also acts as a crude, multi-mode microwave
cavity, is supported by an  
aluminum tube visible in Fig.~\ref{fig:Layers}, panel C. This tube is supported from the downstream end of the central G-10 tube of the refrigerator and serves as a conduit for delivering liquid helium from the run and bypass valves to the sample bath. The mandrel and support tube are efficiently cooled by superfluid creeping from the bath as well as vapor pumped from the bath.

The coils are independently powered and operate at currents of approximately 2~A.  
This design allows a considerable degree of flexibility.  For example, the shims can be used to improve the field homogeneity (or worsen it, to study the effect on the polarization),  or uniformly shift the field to higher or lower values.  It also gives the capability to simultaneously polarize two target samples in opposing directions using a single microwave frequency by shifting the field at one sample above the ESR value and the field at the second below it~\cite{LagerquistPhD}.  We have demonstrated this capability in tests \cite{Maxwell_2018} but have not yet implemented it inside CLAS12.

\begin{figure}[t]
    \centering
    \includegraphics[width=\columnwidth]{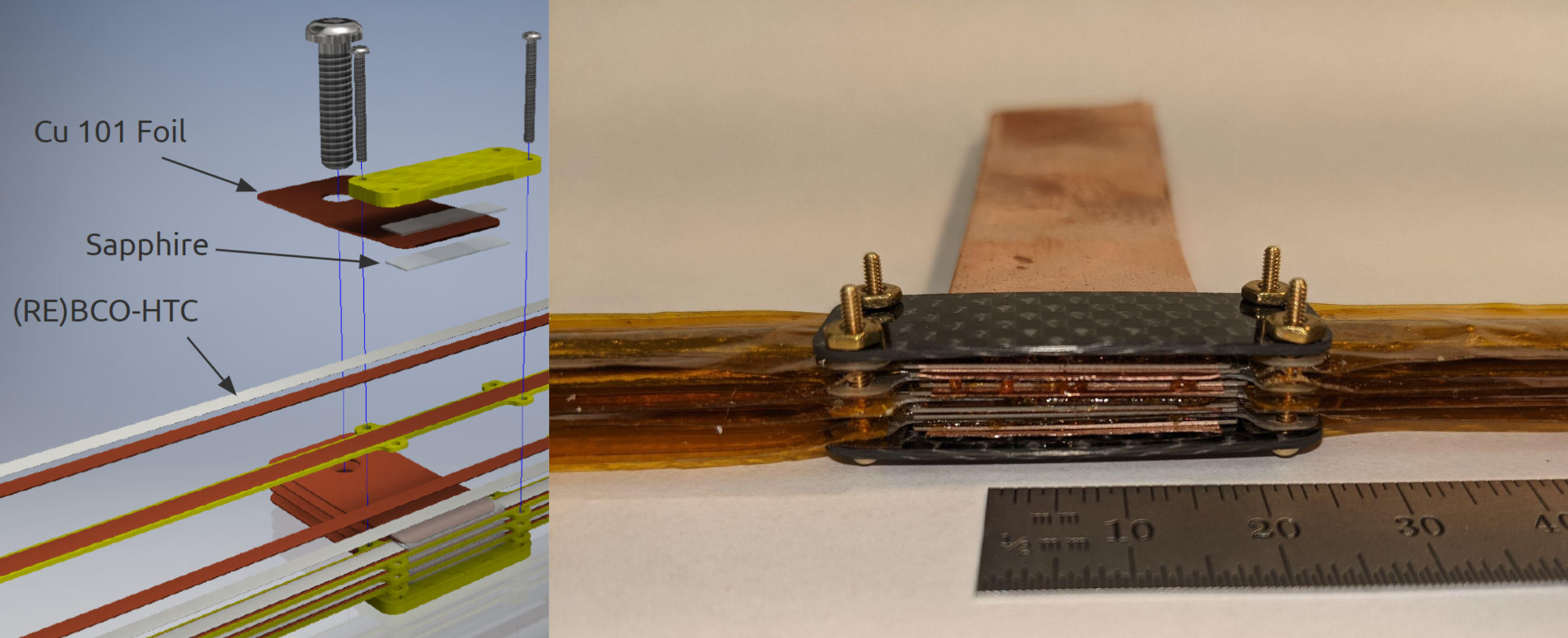}
    \caption{Details of the layered shim coil leads and heat-sinks. Kapton tape is used for electrical insulation.}
    \label{fig:HTC leads assembly}
\end{figure}

The four independent shim coils require eight current leads. These are fabricated in pairs using brass and high-temperature superconducting (ReBCo) ribbons\footnote{SuperPower Inc. SF4050-AP}.  Each pair is soldered to one of four strips cut from doubled-sided, copper-clad printed circuit board material to provide mechanical stability.  
The four strips are stacked together and extend from the warm baffles to the liquid-vapor heat exchanger.   
The layered stack is heat-sunk at three points: at the upstream and downstream ends of the warm baffle, and at the separator. C101 foils interlaced between $0.5$~mm thick sapphire wafers provide thermal contact while maintaining electrical isolation between adjacent leads and the refrigerator baffles, see Fig.~\ref{fig:HTC leads assembly}.

\subsection{Pumping Tube}
\label{sec:PumpingTube}
All components of the 1~K refrigerator are contained within the vacuum-insulated pumping tube that constrains the low-pressure helium vapor pumped from the sample bath.
As shown in panel D of Fig.~\ref{fig:Layers}, the tube is comprised of alternating sections of 
G-10 and aluminum, joined by DP190 gray epoxy.
Moving from the room temperature end to colder end: a modified aluminum ISO 320 flange is joined to a G-10 fiberglass composite tube. The tube is 10-inch inside diameter with a 0.25-inch wall thickness. A thin stainless steel foil is laminated inside the G-10 layers to reduce the permeation of helium gas into the surrounding vacuum. This section is followed by the conical aluminum heat sink for the thermal shield described in Sec.~\ref{sec:Baffles}.    The heat sink reduces the pumping tube's inner diameter down to join a 7.5-inch G-10 composite tube, again with an internally laminated stainless steel foil, and capped with a custom aluminum flange. The custom flange has a V-groove seal \cite{Roth_1994} and a 
127~\textmu m thick polyimide gasket permitting the mating flange to be removed for servicing.  This last flanged section is also aluminum with a 2.75-inch inside diameter. A hemispherical end and 30~mm diameter, 70~\textmu m thick beam window completes the pumping tube.

\subsection{Thermal Shield}
\label{sec:ThermalShield}
A thermal shield, fabricated from annealed copper C101 and aluminum Al6063, surrounds the pumping tube and is clamped to the conical heat sink via a matching conical bolt ring (Fig.~\ref{fig:Layers}, panel F).  Apiezon N vacuum grease is applied to the mating surfaces ensuring good thermal contact between the shield and heat sink of the pumping tube.
Three small G-10 stand-offs epoxied outside of the G-10 pump tube center the heat shield with point contact.  The portion of the shield surrounding the 1~K bath and target sample is made of 0.5~mm thick Al6063 to reduce the energy loss of particles scattered from the target.  The thermal shield is wrapped in 20 layers of multi-layer insulation (MLI), aluminized-mylar/polyester insulation, and instrumented with platinum thermometers to measure its temperature in four locations.  A 20~SLPM flow of helium gas from the separator, pumped through the conical heat exchanger, is sufficient to cool the thermal shield below 20~K and reduces the radiative heat load on the refrigerator from about 15~W to less than 1~mW.

\subsection{Outer Vacuum Chamber}
\label{sec:OuterVacuumChamber}
The outer vacuum chamber for the target cryostat is shown in panel G of Fig.~\ref{fig:Layers} and comprises three sections joined by bolted flanges.  The first two sections are 316L stainless steel, while the downstream section is aluminum 6061 with a CFRP section surrounding the target sample. Carbon fiber was used, due to its superior strength-to-weight ratio, to construct a low density, 1~mm thick shell with low atomic number capable of supporting the vacuum load. This both reduces the energy loss of particles scattering from the target and maximizes the inner diameter
needed to accommodate the essential components of the refrigerator.
It was fabricated in-house as a single layer of 6k double-bias woven sleeve over a double layer of axial-wound 3k tow. A 20~\textmu m aluminum foil inner liner was incorporated into the lamination process to reduce the permeation of air into the vacuum chamber.    
A 32~mm hole covered by a 70~\textmu m thick foil of AL1100 is the exit window for the electron beam. 

\section{Target Bath Assembly, DNP Insert, and Loadlock}
\label{sec:TargetBath}
\subsection{Bath Assembly}
Target samples are cooled to 1~K by placing them in a small, rectangular bath that is filled with superfluid helium from the run or bypass valves.  As described in Sec.~\ref{sec:Introduction}, the NH$_3$ and ND$_3$ samples are removed after being in the beam for 2-3 days due to radiation damage. Unpolarized samples of carbon and polyethylene are also used during the experiment to better understand and quantify the scattering of electrons from nuclei other than hydrogen in the polarized samples.  These are often in the electron beam for one day or less.

\begin{figure*}
    \centering
    \includegraphics[width=0.8\textwidth]{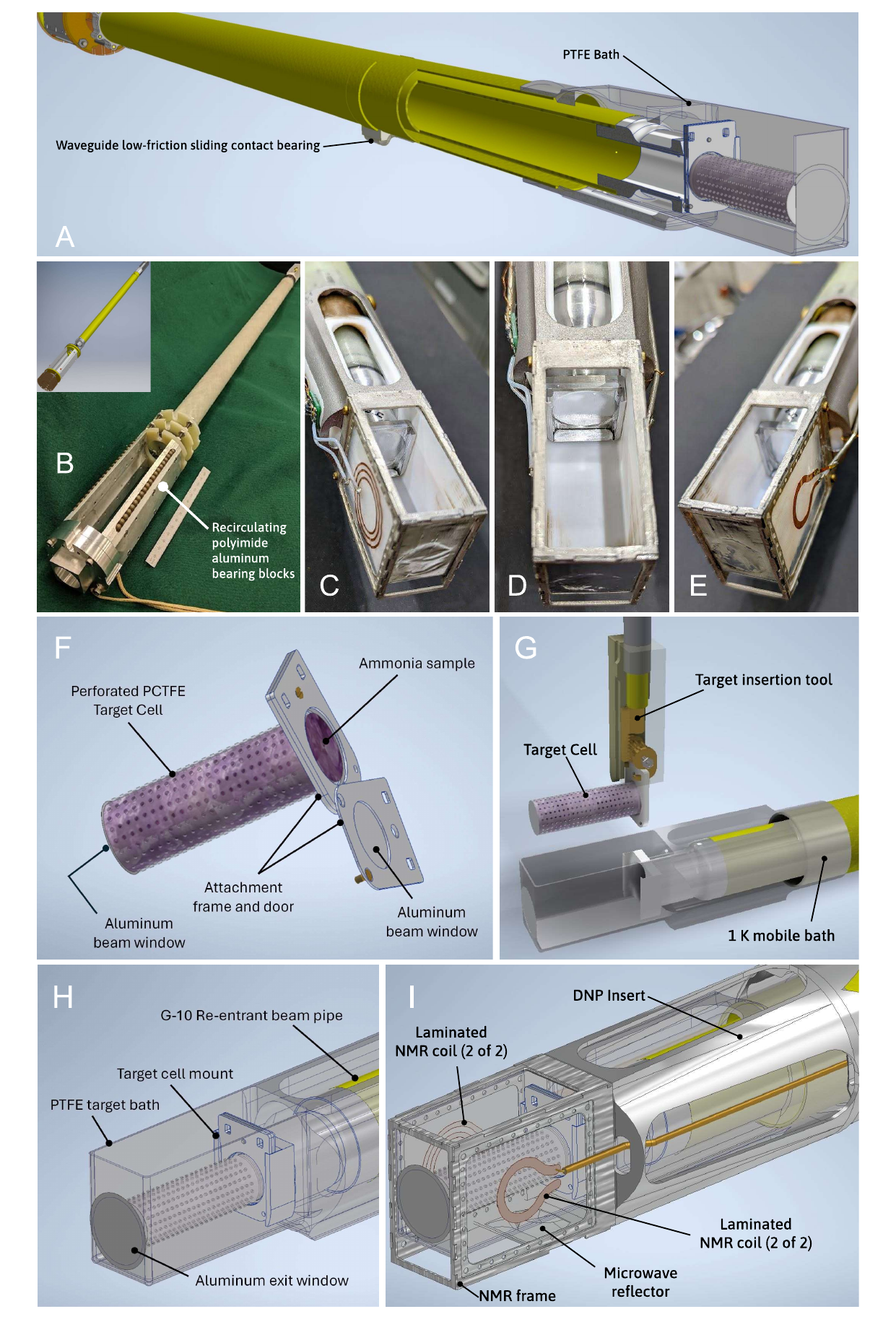}\\
    \caption{Annotated renderings of the sample bath assembly and cell.  A: the sample bath and re-entrant beam line, B: Recirculating polyamide-imide bearings blocks, B inset: Overall bath assembly, C: Deuteron coil and NMR Cold Tank Circuit, D: Bath entrance and exit windows, E: Proton coil, F: Sample Cell, G: Insertion/extraction tool with positive engagement hooks, H: Bath with target sample installed, I  and the frame for the NMR coils. All aluminum windows are 19~\textmu m thick.}
    \label{fig:Trolley}
\end{figure*}

To minimize the schedule impact of frequent changes, we devised a method for rapidly replacing target samples inside the refrigerator to reduce experimental downtime and eliminate the disturbance to other beamline elements such as focusing magnets and beam-position monitors.  With this method, a sample can be removed from the 1~K bath, replaced with a new sample, and cooled again to 1~K in less than 30 minutes, as seen in Fig.~\ref{fig:TargetExchange}. We accomplish this using a bath assembly (Fig.~\ref{fig:Trolley}) that can be retracted approximately 2.5~m from its in-beam position to the room-temperature end
of the cryostat where the target sample can be accessed via a 100~mm diameter load-lock port.  
A similar method was employed for a polarized target at the Stanford Linear Accelerator Center (SLAC) \cite{Ash_1976}.

The assembly consists of a rectangular, thin-walled PTFE sample bath, two concentric G-10 tubes, and a set of recirculating polyamide-imide bearings housed in an aluminum race. (Fig.~\ref{fig:Trolley}, panels A through E)  
The outer tube, approximately 31~mm in diameter and 1~m in length, connects the bath to the aluminum bearing blocks and---along with the sample bath---fills with liquid helium.  This increased volume of liquid helium dampens liquid level fluctuations, making level control more precise. A 27~mm diameter inner tube serves as a re-entrant beam pipe for the electron beam, reducing the quantity of liquid helium in the beam path.  The downstream end of this tube terminates in a 19~\textmu m thick aluminum, Al1235-O, beam window and incorporates a mounting structure for locating and securing target samples inside the bath.  

The target bath has the dimensions 25 x 37 x 69~mm with 0.5~mm thick walls. It is machined from PTFE, which is transparent to both the 140~GHz microwaves for dynamic polarization and the RF used for NMR.  Additionally, neither PTFE nor PFA contain any free hydrogen nuclei that would introduce a false proton NMR signal.  However, PTFE is prone to embrittlement when exposed to high levels of ionizing radiation, and so a 25~mm hole was cut in the downstream end of the bath and covered with a 19~\textmu m  aluminum foil.  Here again, the DP190 two-part epoxy was used, and the PTFE was chemically etched\footnote{FluoroEtch\textregistered Safety Solvent, Acton Technologies Inc.} to improve adhesion. 

\begin{figure}
    \centering
    \includegraphics[width=\columnwidth]{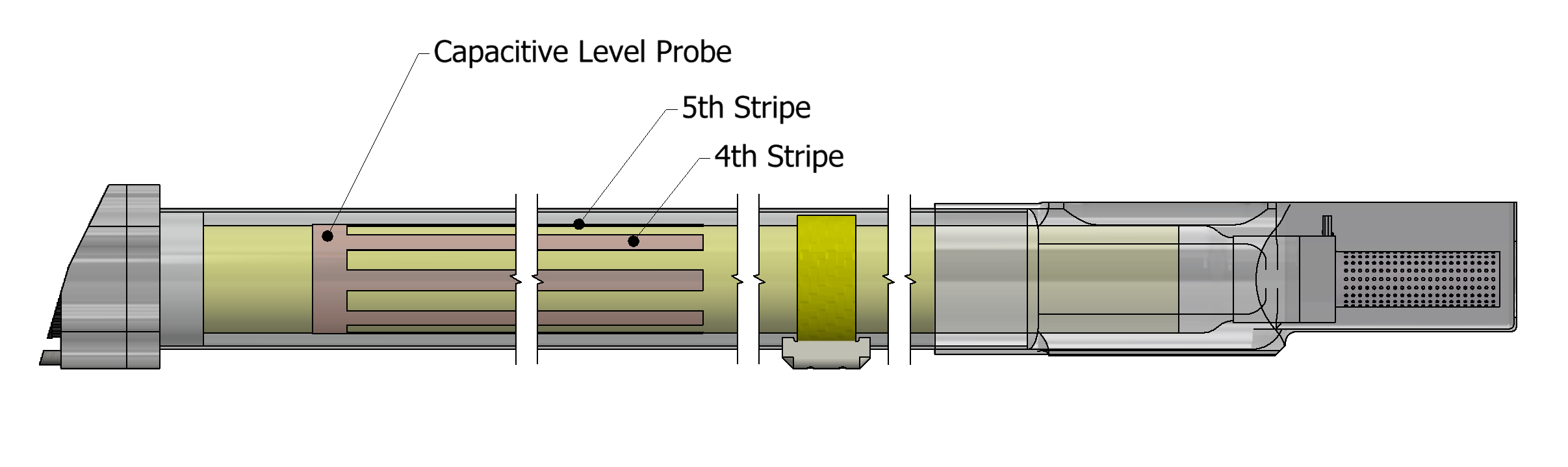}\\
    \includegraphics[width=\columnwidth]{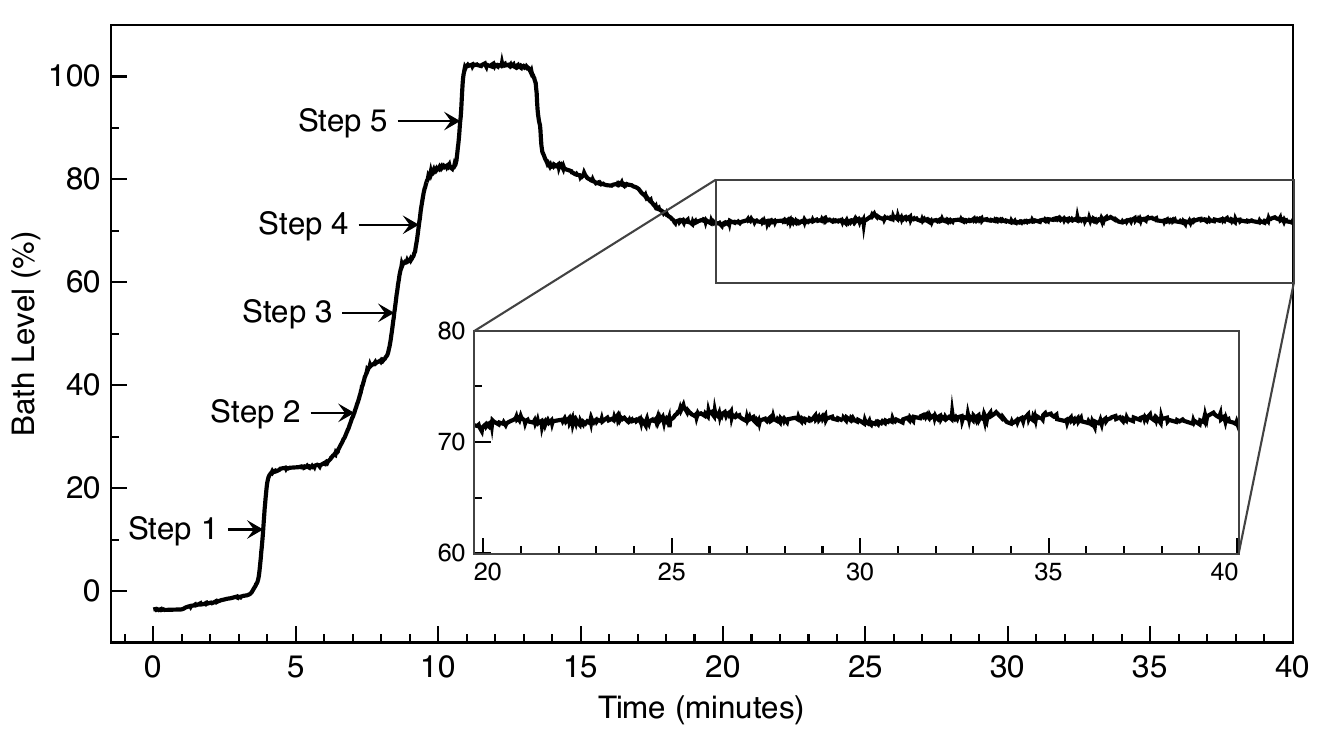}
    \caption{Top: The capacitive level probe for the target bath. Bottom: Filling of the 37~mm deep superfluid bath over the course of ten minutes.   Inset: after twenty minutes the bath stabilizes at a level of (72.0 $\pm$ 0.25)\%, where 1\% corresponds to approximately 0.4 mm.}
    \label{fig:BathPlot}
\end{figure}

Instrumentation in the bath includes two thermometers, a heater of nichrome wire, and a capacitive level probe. These are located slightly upstream from the target sample in the annular space between the inner and outer G-10 tubes of the bath assembly. The probe consists of two parallel plates with a gap of 1 mm.  Each plate has five horizontal copper strips 5~mm wide and separated by 5~mm.  The capacitance, measured by an AC bridge circuit and lock-in amplifier, increases in a stair-step manner as the bath fills with liquid helium (Fig.~\ref{fig:BathPlot}) and provides an unambiguous gauge of the liquid level within the bath. 

\subsection{Bath Motion}
The bath assembly is guided through the central G-10 tube of the cryostat by the DNP insert (Fig.~\ref{fig:Layers}, panel A).  This extends the full length of the cryostat and is removable for servicing. It includes the microwave waveguide, NMR components, and the motion mechanism for the 1~K bath. An over-sized, 6~mm circular waveguide and four cryogenic NMR UT85 cables run along the bottom of the insert and maintain the orientation of the bath as it travels between its in-beam and extraction locations.  The waveguide terminates at the in-beam position inside a 3D printed Direct Laser Metal Sintered (DLMS) aluminum structure that constrains the target bath position, supports NMR components, and incorporates a reflector, distributing the microwave power upwards and onto the target sample (Fig.~\ref{fig:Trolley}, panel I).

\begin{figure}[t]
    \centering
    \includegraphics[width=\columnwidth]{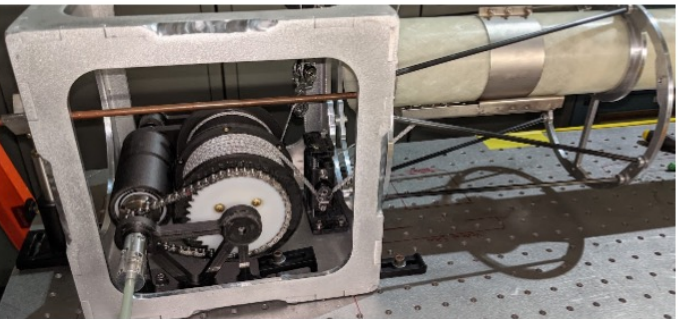}
    \caption{Chain-driven spooling mechanism for the target bath.  The copper portion of the microwave waveguide is visible above the spool,}
    \label{fig:DNP Insert}
\end{figure}

Motion of the target bath assembly is accomplished using two Dyneema\textsuperscript{{\textregistered}} tethers, counter wound onto a single spooling mechanism located at the far upstream end of the DNP insert and target cryostat (Fig.~\ref{fig:Layers} B, and Fig.~\ref{fig:DNP Insert}).  The first tether attaches to the upstream end of the 1~K bath assembly and retracts the bath to the sample exchange position when the spool is rotated counter-clockwise. The second is routed around a Torlon bearing block near the downstream end of the DNP insert and pulls the bath to the in-beam position when the spool is rotated clockwise.  To avoid entanglement, a tensioner is used to maintain the proper lay of the tethers as they coil around the spool, which advances axially upon rotation. 

The tether used to retract the bath acts as a protective sheath for twelve instrumentation wires leading to the target bath. The wires are divided into four bundles and encased in PTFE liners.\footnote{Fluorostore PTFE Spaghetti Tubing AWG 16 Light Wall - FT019047-LW} The bundles are spiraled clockwise together with a locking, counter clockwise, half twist per revolution. This is required to ensure the average length of each conductor is maintained as it is guided around the bends of the blocks, tensioner and spool. This tether configuration was tested at room temperature with failure occurring at 2585 cycles. Rapid mechanical failure of the conductors was observed when this prescription was not followed.  An 18-conductor slip ring commutator\footnote{Moog Components Group, SRA-73683-18}, is attached to the spool and provides stable electrical connection between the bath instrumentation and a vacuum feed-thru.

The low-temperature end of the DNP insert terminates in an open, rectangular frame made from magnesium alloy AZ31B (Fig.~\ref{fig:Trolley}, panels C, D, E, and I).  The frame serves both as a positive stop for the target bath and as an attachment point for two NMR coils that measure the target polarization.
Each coil is cut from thin copper foil and laminated to the beam-left or -right side of the frame using two sheets of a melt-processable fluoropolymer (perfluoroalkoxy, PFA).  One coil is tuned to the 5~T resonance frequency of protons (213~MHz), and the second to that of deuterons (32~MHz).

\subsection{Loadlock}
\begin{figure}
    \centering
    \includegraphics[width=\columnwidth]{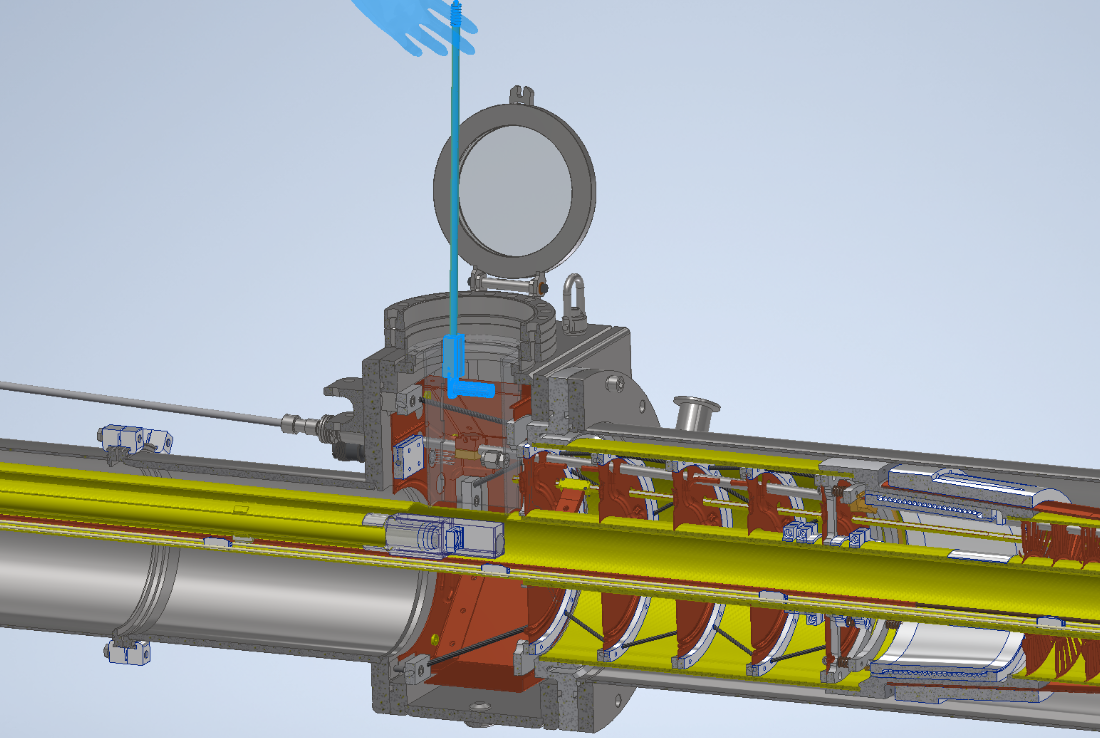}
    \caption{Rendering of the loadlock for exchanging target samples. The target bath is fully retracted under the loadlock, and the sample cell and the loading tool are highlighted in blue, being lowered into the bath.}
    \label{fig:LoadLock}
\end{figure}

Target samples are placed into and extracted from the bath using a bespoke tool that grasps the sample by its aluminum window frame. 
The samples are loaded into the refrigerator through a 6-inch diameter loadlock door with a quartz window. This is mounted to the top of a rectangular, stainless steel box at the mid upstream end of the cryostat (Fig.~\ref{fig:LoadLock}).  The interior of the box is lined with a copper thermal shield that is cooled by the helium vapor pumped from the separator.  It is intended to reduce the heat load on the sample bath during target exchanges.  However, the bath thermometers are located several centimeters upstream from the sample and are outside this thermal shield when the samples are extracted.  Thus located, they may over-estimate the sample temperature during the exchange (Fig.~\ref{fig:TargetExchange}).  A polycarbonate flue is also installed inside the box and serves two purposes.  First, it increases the gas velocity of the helium purge gas which reduces the influx of air into the cryostat during sample exchanges. Second, it helps guide the sample into the bath.

\section{Operation and Performance}
\label{sec:Performance}
\subsection{Cooling from Room Temperature} 
\label{sec:Cooling}
Cooling the refrigerator from room temperature takes approximately 4~h and requires little user intervention.
Prior to cooling, the outer vacuum can is evacuated to a pressure of about 100~\textmu Pa using a turbomolecular pump.  The liquid helium transfer line is inserted between the 500~L dewar and phase separator, and the siphon is initiated by the diaphragm pump.  The mass flow controller at the inlet to this pump maintains the gas flow from the separator at a constant value of about 60 SLPM.  Both the Run and Bypass needle valves are opened, and their positions controlled by PID loops.  The Run valve is controlled by the bath level meter, with a typical set point of 70--80\%.  The Bypass valve is controlled by the mass flow meter at the exhaust of the roots pumps, with a set point of 60~SLPM.  The combined flow though the separator and roots pumps, 120~SLPM, is constant throughout most of the cool down and corresponds to a liquid helium consumption rate of about 10 liters per hour.
As the refrigerator components cool, the Bypass valve slowly closes to maintain
the desired flow rate.  At the same time, flow through the Run valve increases, and the final minutes of the cool down occur with the Bypass valve fully closed and its PID loop disabled.  As the bath begins to fill with liquid, the Run valve automatically adjusts to maintain the liquid level at the specified PID value.  Finally, the mass flow controller for the separator pump is lowered to its nominal value of 20--40~SLPM.

\begin{figure}[t]
    \centering
    \includegraphics[width=\columnwidth]{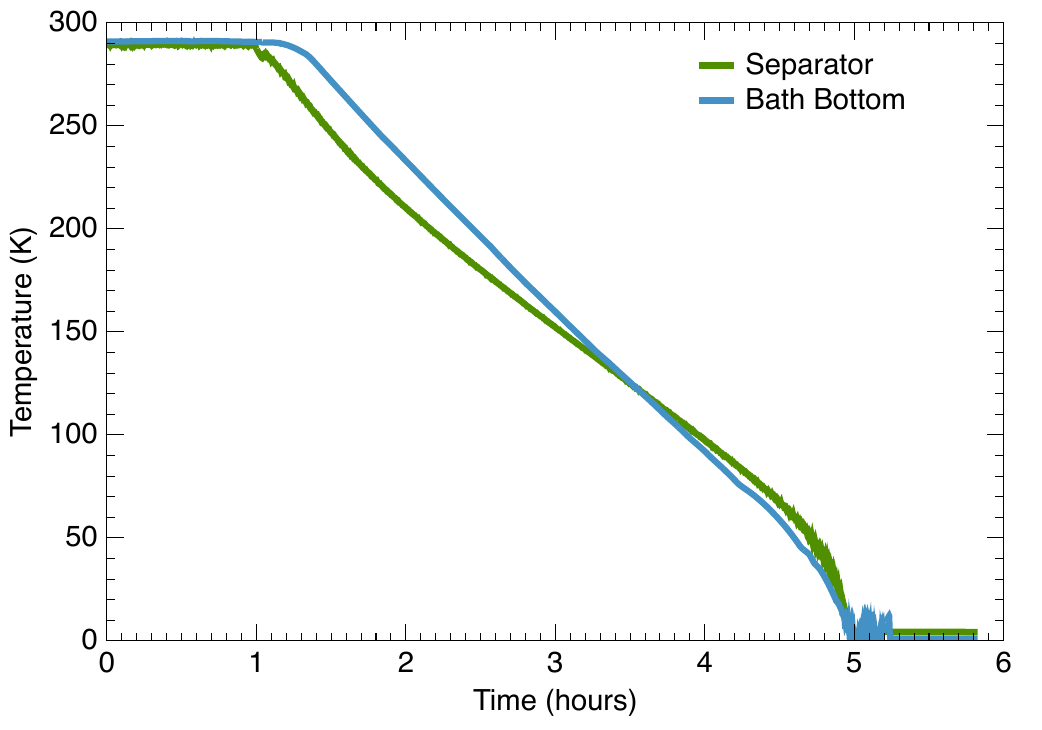}
    \caption{Cooling of the refrigerator from room temperature to 1~K.  The process begins around the 1~hour mark and requires about four hours to complete.  The rapid oscillations at the 5 hour mark are caused by flash vaporization when liquid helium is first produced at the Joule-Thomson valve. }
    \label{fig:Cooldown}
\end{figure}

\subsection{Standard Operation} 
\label{sec:Operation}
\begin{figure*}
    \centering
    \includegraphics[width=0.9\textwidth]{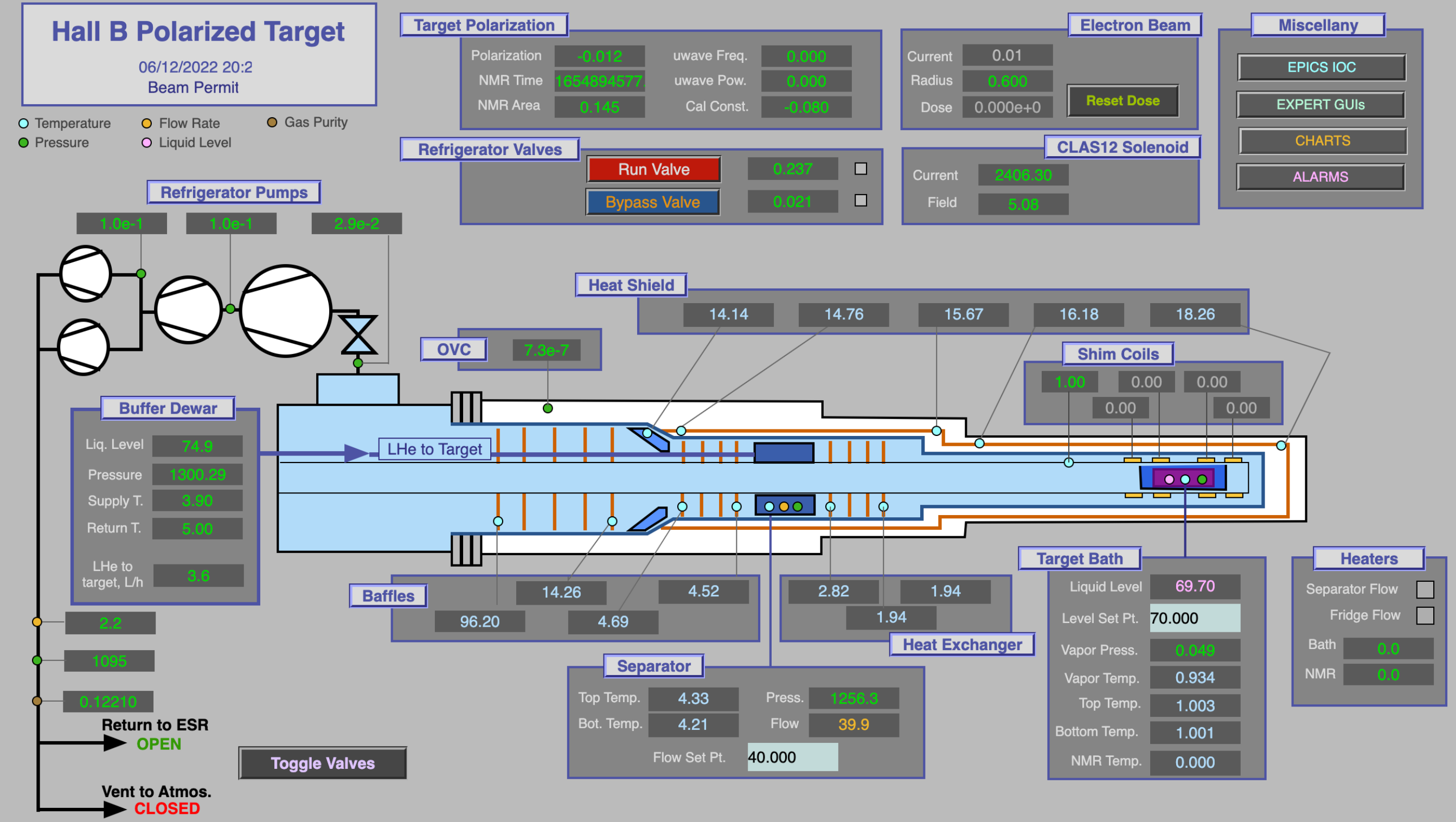}
    \caption{A snapshot of the graphical user interface for controlling and monitoring the 1~K refrigerator that shows operation at its base temperature inside CLAS12.  Sensor locations are marked by small colored circles, and the color code is indicated in the upper left corner.  Sensor units: Temperature (K), Flow (SLPM), Gas Purity (ppm), Liquid Level (\%), and pressure (mbar).  The $^4$He vapor pressure measurement is displayed in torr.}
    \label{fig:EPICS}
\end{figure*}

Monitoring and control of all refrigerator instrumentation is accomplished using the EPICS distributed software control package \cite{White_epics}, which is widely utilized at Jefferson Lab and other large laboratories around the world.
The control screen for the refrigerator operating in Hall B is shown in Fig.~\ref{fig:EPICS}.  In this instance,
the CLAS12 solenoid is energized to 5~T, the microwaves for polarizing the target are off, the bath level is controlled at 70\%, and the refrigerator is operating at a temperature of 0.93~K, as determined by the $^4$He vapor pressure.  This is measured using a room-temperature capacitance manometer connected to a tube that extends beyond the liquid-to-vapor heat exchanger.  The two thermometers in the target bath register a warmer temperature of 1.00~K due to magnetoresistance shifts.  At zero field these thermometers agree with the vapor pressure measurement to within $\pm 0.01$~K.  The evaporation rate from the bath, as measured by the flow meter at the exhaust of the roots pumping system, is 2.2~SLPM, or 6.4 mg/s. 
A thermometer measures a liquid temperature of 1.94~K prior to expansion, so approximately 10\% of the mass flow is due to flash vaporization (see Fig.~\ref{fig:FlashVapor}).  The evaporation rate is observed to depend on the bath level, possibly due to superfluid film creep.  It drops below 2~SLPM when the level is less than 50\% and rises above 6~SLPM when the level exceeds 80\%.

\begin{figure}[t!]
    \centering
    \includegraphics[width=\columnwidth]{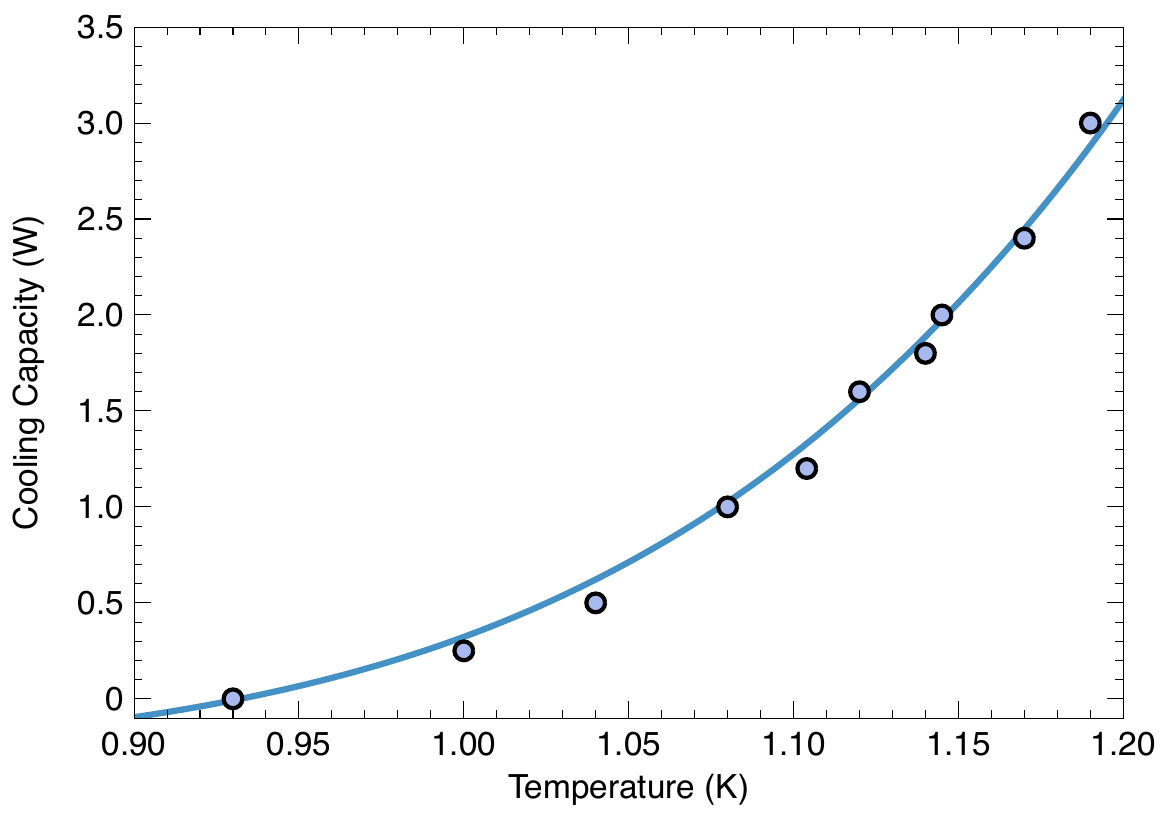}
    \caption{Measured cooling capacity of the refrigerator.  See text for details.}
    \label{fig:CoolingPower}
\end{figure}

The cooling capacity of the refrigerator can be measured using the nichrome heater in the bath. Results, taken with the bath level maintained at 80\%, are shown in Fig.~\ref{fig:CoolingPower}. The blue curve is a calculation of the evaporative cooling process using the latent heat of $^4$He~\cite{Donnelly_1998} and assuming a fixed pumping speed of 4550~m$^3$/h.  It is corrected for 10\% flash evaporation and a no-load evaporation rate from the bath of 6~SLPM.

\begin{figure}[t]
    \centering
    \includegraphics[width=\columnwidth]{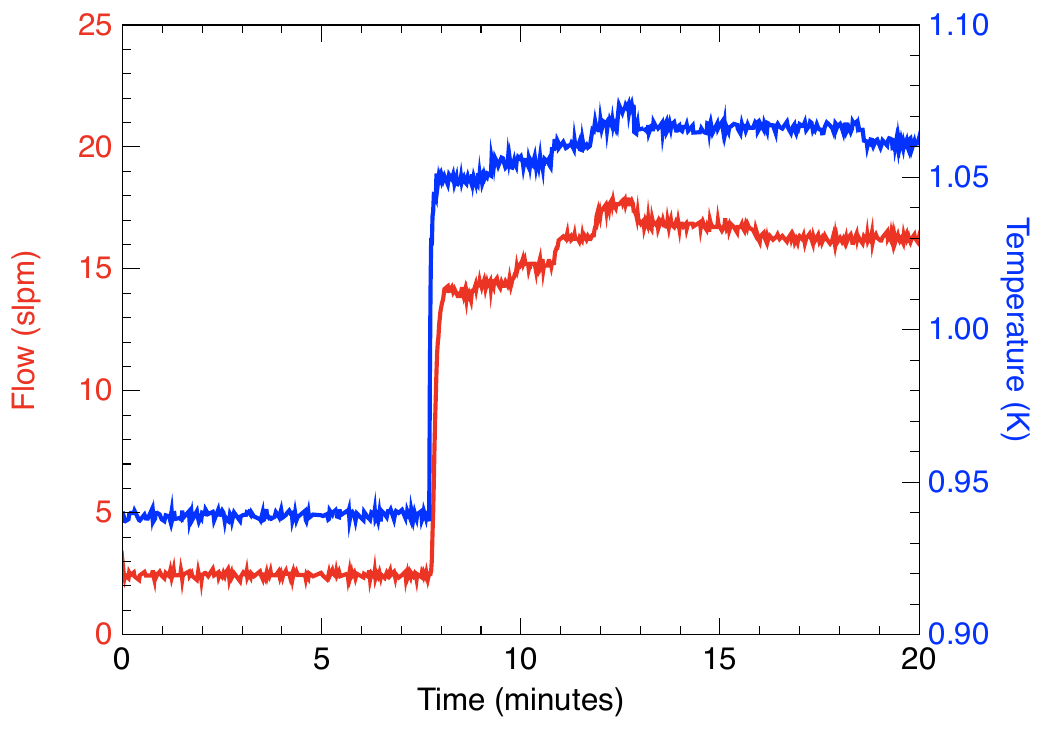}
    \caption{Temperature (blue) and flow (red) response to energizing the microwave generator, which produces a heat load of approximately 0.8~W. The small steps that occur between 8 and 13 minutes are caused by the Run valve adjusting
    to maintain the bath level at 75\%.}
    \label{fig:MicrowavesOnOff}
\end{figure}

The microwave power delivered to the target for maximum polarization is typically 0.5--1~W and can be estimated from the measured evaporation flow rate.  An example is shown in Fig.~\ref{fig:MicrowavesOnOff}, where the refrigerator flow is seen to increase from approximately 2.5 to 16~SLPM when the microwave generator is energized.  At the same time, the bath temperature is observed to warm from 0.94 to 1.06~K.  This corresponds to a heat load of approximately 0.8~W, not adjusted for flash vaporization.  The electron beam deposits an additional head load of 50--100~mW, depending on the beam current.

Stable control of the bath level is critical to the
operation of the polarized target because the space between the top of the target sample and top of the bath is only 15~mm.  
If the level falls below the top of the sample, that portion warms and loses polarization very quickly.  On the other hand, the evaporation rate rises dramatically if the bath overfills, warming the refrigerator above 1~K, with a corresponding drop in target polarization.  Using the capacitive level probe and AC bridge, the Run valve can control the bath level with a precision of about 0.2~mm (Fig.~\ref{fig:BathPlot} inset).

\subsection{Sample Exchange} 
\label{sec:SampleExchange}
Through the course of the Run Group C experiments,
the target sample was exchanged a total of 75 times.  
This section outlines the optimized procedure, refined over the initial stages of the experiment, which requires a minimum of two experienced target personnel to execute.

The exchange begins by preparing the new sample for insertion into the refrigerator.  For samples such as carbon or polyethylene,
no special precautions are needed, but the solid ammonia samples (NH$_3$ and ND$_3$) have to be kept
below 100~K at all times to prevent the loss of the paramagnetic radicals necessary for DNP.
With this in mind, the ammonia samples are stored until needed in Dewars of liquid nitrogen or argon. 
After they are removed from the storage Dewars, any residual liquid is removed by ``drying'' the samples in a \textit{bain-marie} style liquid argon container.  A coil of copper tubing submerged in the lower part of the container cools a stream of nitrogen gas to about 90~K before it flows over the target sample, located in the upper part.  A thermocouple monitors the sample's temperature during this process, which takes about 5~minutes.

\begin{figure}
    \centering
    \includegraphics[width=\columnwidth]{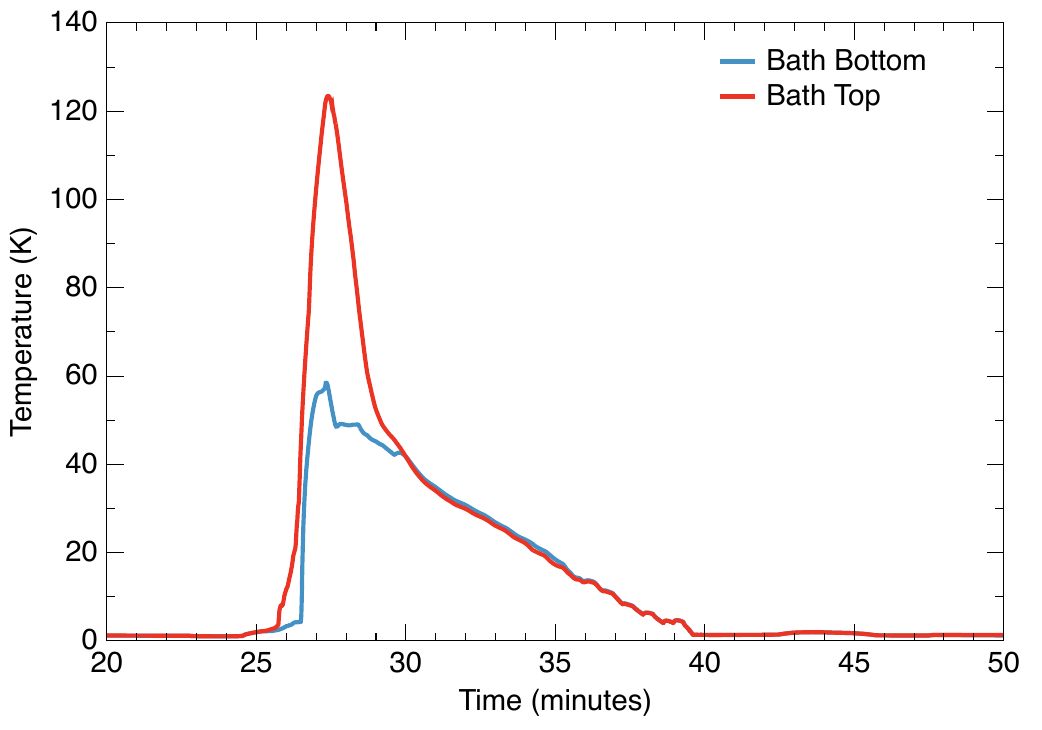}
    \caption{Temperature of the bath thermometers during a target exchange. The procedure begins at the 25 minute mark, and the bath is cooled to 1~K after approximately 15 minutes.  Both thermometers are positioned several centimeters upstream from the target sample and may not accurately reflect its true temperature during the process.}
    \label{fig:TargetExchange}
\end{figure}

While this ``drying'' period progresses, the target bath is retracted approximately 1.2~m to the location of the phase separator,
the gate valve at the inlet to the refrigerator pumps is
the Roots pumps are turned off, leaving the rotary vane pumps running.
The Run and Bypass valves are then fully opened to bleed the pump tube up with cold helium gas.  When the pressure reaches approximately 1/2~atm, the bath is retracted another 1~m
until it is visible through the 6-inch loadlock door.  
When the pump tube reaches atmospheric pressure, the door is opened under a strong purge of cold helium gas and the old sample is replaced with a new one using the
sample insertion tool shown in Fig.~\ref{fig:Trolley}, panel G.  
After the loadlock door is closed and secured, 
the bath is rolled back its in-beam position
as rapidly as possible, and the gate valve is immediately
opened to evacuate the refrigerator pumping tube with the
rotary vane pumps.  The Roots pumps are started when their inlet pressures are sufficiently low.  The Run and Bypass valves are placed in the feedback control states described in Sec.~\ref{sec:Cooling}, and the bath cools and fills in about 15~minutes.  With practice, a target exchange can be performed in less than 30~minutes. This is illustrated in a plot of the bath thermometry during a target sample change shown in Fig.~\ref{fig:TargetExchange}, where a carbon target is replaced by NH$_3$. The actual removal and insertion occurs in under a minute just prior to the peak temperature.  

\section{Summary}
\label{sec:Summary}
This new, horizontal, 1~K refrigerator system was utilized successfully during the Hall B experimental run from 2022 to 2023, cooling dynamically polarized target samples for scattering experiments that examine the spin structure of protons and neutrons at Jefferson Lab. The refrigerator was specifically designed for use inside the CLAS12 detector, which places very severe constraints on its geometry and operation.  The result is a 4~m long cryostat with a novel loadlock system for inserting and extracting target samples into and from the center of the particle detector while maintaining their temperature below 100~K at all times. A base temperature of 0.93~K is achieved with no applied heat load.  The combined heating from the electron beam and microwaves used for dynamic polarization is approximately 1~W, which warms the sample to 1.08~K.  If necessary, additional cooling could be obtained using a Roots pumping system with larger capacity.

During its first use, the target was successfully operated in a high intensity, 11~GeV electron beam for 88 days.  A total of 75 target exchanges were performed over this period, without incident.  The method employed here eliminated the need to disassemble and replace large sections of the electron beam line each time the target sample was replaced and saved an estimated eighteen days of valuable beam time.


\section*{Acknowledgements}
The authors thank Xiangdong Wei (Jefferson Lab), Derek Holmberg (College of William \& Mary), Zulkaida Akbar (University of Virgina), Sebastian Kuhn (Old Dominion University), and Susan Schadmand (Institut f{\" u}r Kernphysik, J{\" u}lich) for their assistance during the Run Group C experiments.  We also thank members of the Hall B engineering and technical staff, the JLab Survey and Alignment Group, and the JLab Radiation Control Group for their support during the experiments.   The authors remember D. Crabb and M. Seely, whose work in polarized targets inspired and instructed us and many others. The irradiated ammonia samples were kindly provided by the University of Virginia Polarized Target Group.  This work is supported by the U.S. Department of Energy, Office of Science, and Office of Nuclear Physics, under contracts DE-AC05-06OR23177 and DE-FG02-96ER40960. The target project was partially supported by an NSF-MRI grant awarded in 2009.

\bibliographystyle{elsarticle-num} 
\bibliography{bibliography}

\begin{thebibliography}{10}
\expandafter\ifx\csname url\endcsname\relax
  \def\url#1{\texttt{#1}}\fi
\expandafter\ifx\csname urlprefix\endcsname\relax\def\urlprefix{URL }\fi
\expandafter\ifx\csname href\endcsname\relax
  \def\href#1#2{#2} \def\path#1{#1}\fi

\bibitem{Abragam_1962}
A.~Abragam, M.~Borghini, P.~Catillon, J.~Coustham, P.~Roubeau, J.~Thirion, Diffusion de protons polarizes de 20 mev par une cible de protons polarizes et mesure preliminaire de parameter $c_{nn}$, Phys. Lett. 2 (1962) 310--311.

\bibitem{Chamberlain_1963}
O.~Chamberlain, C.~Jeffries, C.~Schultz, G.~Shapiro, L.~{Van Rossum}, Pion scattering from a polarized target, Physics Letters 7~(4) (1963) 293--295.
\newblock \href {https://doi.org/https://doi.org/10.1016/0031-9163(63)90338-X} {\path{doi:https://doi.org/10.1016/0031-9163(63)90338-X}}.

\bibitem{CLAS12}
V.~D. Burkert, et~al., {The CLAS12 Spectrometer at Jefferson Laboratory}, Nucl. Instrum. Meth. A 959 (2020) 163419.
\newblock \href {https://doi.org/10.1016/j.nima.2020.163419} {\path{doi:10.1016/j.nima.2020.163419}}.

\bibitem{Fair_2020}
R.~Fair, et~al., {The CLAS12 superconducting magnets}, Nucl. Instrum. Meth. A 962 (2020) 163578.
\newblock \href {https://doi.org/10.1016/j.nima.2020.163578} {\path{doi:10.1016/j.nima.2020.163578}}.

\bibitem{Maly_2008}
T.~Maly, G.~T. Debelouchina, V.~S. Bajaj, K.-N. Hu, C.-G. Joo, M.~L. Mak–Jurkauskas, J.~R. Sirigiri, P.~C.~A. van~der Wel, J.~Herzfeld, R.~J. Temkin, R.~G. Griffin, Dynamic nuclear polarization at high magnetic fields, The Journal of Chemical Physics 128~(5) (2008) 052211.
\newblock \href {https://doi.org/10.1063/1.2833582} {\path{doi:10.1063/1.2833582}}.

\bibitem{Goertz_2002}
S.~Goertz, W.~Meyer, G.~Reicherz, {Polarized H, D and He-3 targets for particle physics experiments}, Prog. Part. Nucl. Phys. 49 (2002) 403--489, [Erratum: Prog.Part.Nucl.Phys. 51, 309--312 (2003)].
\newblock \href {https://doi.org/10.1016/S0146-6410(02)00159-X} {\path{doi:10.1016/S0146-6410(02)00159-X}}.

\bibitem{Averett_1999}
T.~D. Averett, et~al., {A Solid polarized target for high luminosity experiments}, Nucl. Instrum. Meth. A 427 (1999) 440--454.
\newblock \href {https://doi.org/10.1016/S0168-9002(98)01431-4} {\path{doi:10.1016/S0168-9002(98)01431-4}}.

\bibitem{Keith_2003}
C.~D. Keith, et~al., {A polarized target for the CLAS detector}, Nucl. Instrum. Meth. A 501 (2003) 327--339.
\newblock \href {https://doi.org/10.1016/S0168-9002(03)00429-7} {\path{doi:10.1016/S0168-9002(03)00429-7}}.

\bibitem{Pierce_2014}
J.~Pierce, et~al., {Dynamically polarized target for the $g^p_2$ and $G^p_E$ experiments at Jefferson Lab}, Nucl. Instrum. Meth. A 738 (2014) 54--60.
\newblock \href {https://doi.org/10.1016/j.nima.2013.12.016} {\path{doi:10.1016/j.nima.2013.12.016}}.

\bibitem{Meyer_2004}
W.~Meyer, {Ammonia as a polarized solid target material: A review}, Nucl. Instrum. Meth. A 526 (2004) 12--21.
\newblock \href {https://doi.org/10.1016/j.nima.2004.03.145} {\path{doi:10.1016/j.nima.2004.03.145}}.

\bibitem{Roubeau_1966}
P.~Roubeau, Horizontal cryostat for polarized proton targets, Cryogenics 6~(4) (1966) 207--212.
\newblock \href {https://doi.org/10.1016/0011-2275(66)90069-5} {\path{doi:10.1016/0011-2275(66)90069-5}}.

\bibitem{Warren_1968}
J.~Warren, W.~H., W.~G. Bader, Superconductivity measurements in solders commonly used for low temperature research, Review of Scientific Instruments 40~(1) (1969) 180--182.
\newblock \href {https://doi.org/10.1063/1.1683729} {\path{doi:10.1063/1.1683729}}.

\bibitem{LagerquistPhD}
V.~Lagerquist, Design and construction of a longitudinally polarized solid nuclear target for clas12, Phd dissertation, Old Dominion University (2023).
\newblock \href {https://doi.org/10.25777/36yz-ft35} {\path{doi:10.25777/36yz-ft35}}.

\bibitem{Maxwell_2018}
J.~Maxwell, {NMR Measurements for Solid Polarized Targets at Jefferson Lab}, PoS SPIN2018 (2019) 102.
\newblock \href {https://doi.org/10.22323/1.346.0102} {\path{doi:10.22323/1.346.0102}}.

\bibitem{Roth_1994}
A.~A. Roth, Vacuum sealing techniques / A. Roth., American Vacuum Society classics, American Institute of Physics, New York, 1994.

\bibitem{Ash_1976}
W.~W. Ash, {SLAC-Yale polarized proton target}, AIP Conf. Proc. 35 (1976) 485--493.
\newblock \href {https://doi.org/10.1063/1.30990} {\path{doi:10.1063/1.30990}}.

\bibitem{White_epics}
K.~S. White, M.~Bickley, W.~Watson, \href{https://www.osti.gov/biblio/754681}{The evolution of jefferson lab's control system}, Tech. rep., Thomas Jefferson National Accelerator Facility (TJNAF), Newport News, VA (United States) (10 1999).
\newblock \href {https://doi.org/10.2172/754681} {\path{doi:10.2172/754681}}.
\newline\urlprefix\url{https://www.osti.gov/biblio/754681}

\bibitem{Donnelly_1998}
R.~J. Donnelly, C.~F. Barenghi, The observed properties of liquid helium at the saturated vapor pressure, Journal of Physical and Chemical Reference Data 27~(6) (1998) 1217--1274.
\newblock \href {https://doi.org/10.1063/1.556028} {\path{doi:10.1063/1.556028}}.

\end{thebibliography}
\end{document}